\documentstyle[pre,aps,preprint,epsf,eqsecnum]{revtex}
\begin{document}
\draft
\title{Fracture Patterns Induced by Desiccation in a Thin Layer}
\author{So Kitsunezaki\footnotemark[1]}
\address{Department of Physics, Nara Women's University, Nara 630-8506, Japan}
\date{\today}
\maketitle
\footnotetext[1]{kitsune@minnie.disney.phys.nara-wu.ac.jp}
\begin{abstract}
We study a theoretical model of mud cracks, that is, the fracture patterns resulting from the contraction with drying in a thin layer of a mixture of granules and water.
In this model, we consider the slip on the bottom of this layer and the relaxation of the elastic field that represents deformation of the layer.
Analysis of the one-dimensional model gives results for the crack size that are consistent with experiments.
We propose an analytical method of estimation for the growth velocity of a simple straight crack to explain the very slow propagation observed in actual experiments.
Numerical simulations reveal the dependence of qualitative nature of the formation of crack patterns on material properties.
\end{abstract}
\pacs{46.35.+z,46.50.+a,47.54.+r,62.20.Mk}

\section{Introduction} \label{Introduction}

Many kinds of mixtures of granules and water, such as clay, contract upon desiccation and form cracks.
These fracture patterns are familiar to us as ordinary mud cracks.
However, the fundamental questions about these phenomena have not yet been answered theoretically.
The problems which need to be addressed include determining the condition under which fragmentation occurs, the dynamics displayed by cracks, and the patterns which grow.

In simple and traditional experiments on mud cracks, a thin layer of a mixture in a rigid container with a horizontal bottom is prepared left to dry at room temperature \cite{Kindle1917,Groisman94,Ito98,Mitsui95,Nishimoto99}.
Typically, clay, soil, flour, granules of magnesium carbonate and alumina are used.
In almost all cases, cracks extend from the surface to the bottom of the layer and propagate horizontally along a line, forming a quasi-two-dimensional structure.   
Typically we observe a tiling pattern composed of rectangular cells in which cracks mainly join in a T-shape.
Groisman and Kaplan carried out more detail experiments with coffee powder and reported 
1) that the size of a crack cell after full drying is nearly proportional to the thickness of the layer and larger in the case of a ``slippery'' bottom,
2) that the velocity of a moving crack is almost independent of time for a given crack and very slow on the order of several millimeters per minute, but that it differs widely from one crack to another, and
3) that as the layer becomes thin, there is a transition to patterns which contain many Y-shape joints and unclosed cells owing to the arrest of cracks\cite{Groisman94}.

Another experimental setup was used by Allain and Limat \cite{Allain95}.
This setup produces cracks that grow directionally by causing evaporation to proceed from one side of the container.
Sasa and Komatsu have proposed a theoretical model for such systems \cite{Sasa97}.

Fragmentation of coating or painting also arises from desiccation,
This has been studied theoretically by some people \cite{Hornig96,Handge97,Leung97}.
From the viewpoint that fractures are caused by slow contraction, these problems can be thought of as belonging to the same category as thermal cracks in glasses \cite{Yuse93,Yuse97,Ronsin97,Hayakawa94a,Hayakawa94b,Sasa94} and the formation of joints in rocks brought by cooling \cite{Holmes78,Wearie83}.
In addition, we note that mixtures of granular matter and fluid have properties that vary greatly from that of complete elastic materials, in particular, dissipation and viscoelasticity.
The propagation of cracks in such media has been investigated recently using developments in nonlinear physics \cite{Fineberg91,Marder96,Barber89,Langer92,Langer93,Ching96a,Ching96b,Holian97,Fukuhara98,Kessler98}.

\

In this paper, we undertake a theoretical investigation of the experiments described above.
We treat such system as consisting of fractures arising from quasi-static and uniform contraction in thin layers of linear elastic material.

In Sec. \ref{1d model-1}, we propose a one-dimensional model.
Our model takes into account the slip displacement on the bottom of a container, because most of the experiments can not be assumed to obey a fixed boundary condition.
We can investigate the development of the size of a crack cell by applying a fragmentation condition to the balanced states of the elastic field. 

In Sec. \ref{1d model-2}, we report the analytical results of our one-dimensional model.
Here we consider both the critical stress condition and the Griffith criterion as the fragmentation condition.
We consider these two alternative criteria because the nature of the breaking condition in mixtures of granules and water is not clear.
The critical stress condition predicts that the final size of a crack cell is proportional to the thickness of the layer and that, in the case of a slippery bottom, it becomes much larger than the thickness.
These predictions seem to be consistent with the experimental results.
In contrast, we find that the Griffith criterion predicts a different relation between the final size of a cell and the thickness.

In Sec. \ref{2d model-1}, we extend the model to two dimensions and investigate the time development of a crack.
In order to describe the relaxation process of the elastic field, we use the Kelvin model while taking into account the effect of the bottom of the container.
We assume the stress, excluding dissipative force, to be constant in the front of a propagating crack tip and evaluate the velocity of a simple straight crack tip analytically.
Our results indicate that cracks advance at very slow speed in comparison with  the sound velocity.

In Sec. \ref{2d model-2}, we report on the numerical simulations of our model that reproduce fracture patterns similar to those in real experiments.  
The growth of the patterns exhibits qualitative differences depending on the elastic constants and the relaxation time.
As the relaxation time becomes smaller, in particular, we observe the growth of fingering patterns with tip splitting rather than side branching of cracks.

Finally, we conclude the paper with a summary of the results and a discussion of the open problems in Sec. \ref{Conclusions}.

\section{Modeling of fracture caused by slow shrinking} \label{1d model-1}

We analyze the formation of cracks induced by desiccation in terms of the following four processes.
\begin{enumerate}
\item The water in a mixture evaporates from the surface of a layer.
\item Each part of the mixture shrinks upon desiccation.
\item Stress increases in the material because contraction is hindered near the bottom of a container.
\item Fracture arises under some fragmentation condition.
\end{enumerate}

In this section, we examine each process individually and construct a one-dimensional model, where we introduce simple assumptions regarding the unclear properties of granular materials.
Some similar models have been proposed previously \cite{Hornig96,Sasa97}.
One-dimensional models assume that cracks are formed one at a time, each propagating along a line and thereby dividing the system into two pieces separated by a boundary with one-dimensional structure.
Using this assumption, we can ignore the propagation of cracks and consider the development of patterns by using only the condition of separation. 

\begin{enumerate}
\item
From a microscopic viewpoint, water either exists in the inside of the particles of granular materials or acts to create bonds between the particles.
Here, we can introduce the water content averaged over a much larger area than that of a single particle and measure the degree of drying.
When the thickness of a layer $H$ is sufficiently thin and the characteristic time of desiccation $T_d$ is very large, the water content in the layer can be considered uniform.
Assuming that water transfers diffusively in a layer, the sufficient condition here is that $H^2/T_d$ is much smaller than the diffusion constant.
Therefore, we restrict our consideration to the case of the uniform water distribution and exclude the process of water transfer from the model.
\item
The main cause of contraction is the shrinking of particles in the mixture arising from desiccation.
The water content is considered to determine the shrinking rate in the case of uniform contraction in which all the boundaries of the mixture are stress-free.
We refer to this shrinking rate as ``free shrinking rate'' in the following discussions and this concept is used in place of the concept of the water 
This makes clear the relation between the present problems and those involving fractures induced by other causes, such as temperature gradient \cite{Yuse93,Yuse97}, with slow contraction.
We note, however, that it is more difficult to measure the free shrinking rate than the water content experimentally and it is thus necessary to know their relation to compare our theory with experiments on the time development of patterns.

 The contraction force is thought to arise from the water bonds among particles.
We estimate the Reynolds number $R_{e}$ to consider the behavior of the water in a bond.
The diameter of a particle $R$ is generally about $0.1mm$, and the kinematic viscosity of water $\nu$ is about $1 mm^2/s$.
Although the propagation of a crack causes the displacement of surrounding particles with opening the crack surfaces, the velocity of the displacement is smaller than the crack speed itself, except in the microscopic region at the crack tip.
Therefore we estimate the typical velocity of water $V$ in the bulk of a mixture to be smaller than the crack speed.
The crack speed has been measured as about 0.1 mm/s in experiments and it is, of course, considerably faster than the shrinking speed of the horizontal boundary with desiccation, which is typically about $10mm/day$.
Thus the Reynolds number $R_{e}=RV/\nu$ is estimated to be smaller than 1/100.
We expect that the water among the particles behaves like a viscous fluid and that the mixture displays strong dissipation. 

If a material displays strong dissipation and shrinks quasi-statically, the elastic field is balanced to minimize the free energy, except during the time when cracks are propagating.
We deal with balanced states for the present and return to the problem of relaxation in order to treat the development of cracks in Sec. \ref{2d model-1}.
Because mixtures of granules and fluids have many unclear properties with respect to elasticity, we idealize them as linear elastic materials.
When the free volume-shrinking rate $C_{_V}$ is uniform, it is well known that the free energy density of a uniform and isotropic linear elastic material is given in   terms of the stress tensor $u_{ij}$ in the form \cite{LandauElasticTheory},
\begin{equation}
 e_{_V}=\frac{1}{2}\kappa_{_V}(u_{ll}+C_{_V})^2+\mu\left(u_{ik}-\frac{1}{3}u_{ll}\delta_{ik}\right)^2,
 \label{free energy}
\end{equation}
where $\kappa_{_V}$ and $\mu$ are the elastic constants and repeated indices indicate summation.
The stress tensor is expressed as
\begin{equation}
 \sigma_{_Vij} \equiv \frac{\partial e_{_V}}{\partial u_{ij}}
 = (\lambda u_{ll}+\kappa_{_V} C_{_V})\delta_{ij}+2\mu u_{ij},
 \label{stress tensor}
\end{equation}
where $\kappa_{_V}\equiv \lambda+2\mu/3$.
As a result, the balanced equation of the elastic field $\partial \sigma_{_Vij}/\partial x_j=0$ does not include the shrinking rate $C_{_V}$ for linear elastic materials with uniform contraction and it is the same as in the case of the elastic materials without shrinking.
The effect of contraction appears only through the boundary conditions.
\item
After the formation of cracks divides the system into cells, each cell is independent of the others, because the vertical surfaces of cracks become stress-free boundaries.
Without considering the boundary conditions on the lateral sides of the container, we can simplify the problem by starting with the initial condition that the system has stress-free boundaries on its lateral sides.
In contrast to the lateral and the top surfaces, the bottom of a layer is not a stress-free boundary.
The difference among the boundary conditions produces strain with contraction and then stress.
This is the cause of fracture.

We observe the slip of layers along the bottom in most experiments.
Thus we introduce slip displacement with a frictional force into the model.
Because the frictional force is caused by the water between the bottom of a layer and the container, it is considered to remain finite even in the limit of vanishing thickness of a layer $H$ \cite{Groisman94}.
In order to understand the effects of friction for crack patterns, we simplify the maximum frictional force per unit area of the surface to be constant and independent of $H$ without making a distinction between static and kinetic frictional effects.
\item
Cracks propagate very slowly in a mixture of granules and water.
Because this propagation resembles quasi-static growth of cracks, the first candidate of the fragmentation condition is the Griffith criterion applied to the free energy of the entire system.

However, we note that ordinary brittle materials break instantaneously, not quasi-statically, in the situation that the stress increases without fixing the deformation of the system.
The situation is similar to that in a shrinking mixture.
We need to consider the possibility that cracks in a mixture propagate slowly owing to dissipation.
Therefore we consider two typical fragmentation conditions, the critical stress condition and the Griffith criterion, in the first and the last halves of Sec.\ref{1d model-2}, respectively. 

In the context of the critical stress condition, the fragmentation condition is that the maximum principal stress exceeds a material constant at breaking.
This has also been used in many numerical models because of the technical advantage of the local condition.
Our model introduces the critical value for the energy density as an equivalent condition.
We note that the energy of the system before the fragmentation is higher than after the fragmentation because the critical value is a material constant independent of the system size.

In contrast, using the Griffith criterion as an alternative fragmentation condition stipulates that the energy changes neither before nor after breaking.
This condition is used by Sasa and Komatsu in their theory \cite{Sasa97}.
\end{enumerate}

With the above considerations, we construct a one-dimensional model, following the lead of Sasa and Komatsu \cite{Sasa97}.
As we show in Fig. \ref{1dmodel.eps}, we consider a chain of springs by a distance $a$ as the discrete model of the thin layer of an elastic material, where we number the nodes $i=-N,-N+1,...,N-1,N$ for a system of half-size $L:=aN$.
In order to represent the vertical direction of a sufficiently thin layer, we introduce vertical springs with length $H$ which connect each node to an element on the bottom.
The vector $(u_i,v_i)$ represents the horizontal and vertical displacements of the $i$th node, and $w_i$ is the horizontal displacement of the element on the bottom connected with the node. 
The shrinking of a material is modeled by decreasing the natural length of the springs.
We assume an isotropic material with a linear free shrinking rate $s$, making the natural lengths of both the horizontal and vertical springs to be $1-s$ times the initial lengths, i.e. $(1-s)a$ and $(1-s)H$.
 
If the vertical springs are simple ordinary springs, linear response is lost under shearing strain.
We therefore add non-simple springs to the vertical direction which produce a horizontal force in the case that $u_i\neq w_i$ to represent a linear elastic material.
This type of spring is used in the model of Hornig et al. \cite{Hornig96}.
The energy of the system is described by
\begin{eqnarray}
E=\frac{1}{2}\sum_{i=-N}^{N-1}K_1(u_{i+1}-u_i+sa)^2\nonumber\\
+\frac{1}{2}\sum_{i=-N}^{N}[K_2(u_i-w_i)^2+K'_2(v_i+sH)^2],
\label{descrete 1d energy}
\end{eqnarray}
where $K_1,K_2$ and $K'_2$ are the spring constants.
Because $v_i$ is included in the last term independently of both $u_i$ and $w_i$, it follows that $v_i=-sH$ in the balanced states, and then this term vanishes.
This indicates that the model neglects the horizontal stress arising from vertical contraction.
Thus all we need to do is minimize the energy (\ref{descrete 1d energy}) without the last term in order to find the horizontal displacements $u_i$ and $w_i$.
In the continuous limit, $a \rightarrow 0$, the above energy should be described using an independent energy density for $H$, as in the case of (\ref{free energy}) for a linear elastic material.
We scale the space length by the thickness of a layer $H$ and introduce the space coordinate $x:=ai/H$.
Through the transformation to non-dimensional variables $L\rightarrow LH$, $u_i \rightarrow H u(x)$, $w_i \rightarrow H w(x)$ and $E \rightarrow H^2E$, the energy (\ref{descrete 1d energy}) becomes
\begin{mathletters}
\label{1d energy}
\begin{eqnarray}
E=\int_{-L}^{L}dx\{e_1(x)+e_2(x)\},\\
e_1(x)=\frac{1}{2}k_1(u_x+s)^2\\
\mbox{and}\hspace{1em}e_2(x)=\frac{1}{2}k_2(u-w)^2,
\end{eqnarray}
\end{mathletters}
and we know that both $k_1:=\frac{a}{H}K_1$ and $k_2:=\frac{H}{a}K_2$ are the independent constants  of $H$.

We introduce the maximum frictional force $F_s$ for the slip of the elements on the bottom, as explained above.
The vertical spring pulls the $i$th element along the bottom with the force $F_i=K_2(u_i-w_i)$.
Each element on the bottom remains stationary if $|F_i|<F_s$, and, if not, it slips to a position at which $|F_i|=F_s$ is satisfied. 
The slip condition is expressed by the energy density of a vertical spring through the following rule in the previous continuous $a \rightarrow 0$ limit: 
\begin{equation}
e_2(x)>\frac{1}{2}k_1 s_s^2
\hspace{1em}\Rightarrow \hspace{1em} w(x)=u(x)\pm \frac{s_s}{q}.
\label{slip condition}
\end{equation}
Here the choice of the sign depends on the direction of the force.
The constants $s_s$ and $q$ are defined by the equations
\begin{equation}
\frac{1}{2}k_1 s_s^2\equiv \frac{F_s^2}{2k_2a^2}\hspace{1em}\mbox{and}
\hspace{1em} q\equiv\sqrt{\frac{k_2}{k_1}}.
\label{q and F_s}
\end{equation}

We note the constant $q$ is order 1, because its square represents a ratio of certain elastic constants which are the same order in ordinary materials.

Neglecting the short periods during which the system experiences cracking and slip, (\ref{1d energy}) and (\ref{slip condition}) constitute the closed form of our one-dimensional model with the fragmentation condition given in the next section.

\section{Analysis of the one-dimensional model}\label{1d model-2}

We here report analytical results of our one-dimensional model for the typical two fragmentation conditions, i.e, the critical stress condition and the Griffith condition.

\subsection{The Critical Stress Condition}

The critical stress condition demands that the maximum principal stress exceeds a material constant at the fragmentation.

In the case of this condition, we can generally demonstrate that it is difficult to treat the bottom surface as a fixed boundary for a uniform and isotropic elastic material.
We first explain it before the analysis of the one-dimensional model.
Let us think of the layer of a linear elastic material contracting with a fixed boundary condition on the bottom.
It is shrinking more near the top surface, and the cross section assumes the form of a trapezoid as we show  schematically in Fig. \ref{boundary.eps}.
We compare the stress at the following three points: (A) the horizontal center of the cell near the bottom; (B) the lateral point near the bottom; and (C) the horizontal center above the bottom.
The horizontal tensions at A and B are the same because of the fixed boundary condition on the bottom.
Although the stress at C is as horizontal as at A, the strength is weaker.
B is also pulled in the direction along the lateral surface due to the deformation.
Using A, B, and C to represent the respective strengths of the maximum principle stresses at these three points, we find that they are related as $\mbox{C}< \mbox{A} < \mbox{B}$, and we expect generally that fracture arises at B before either A or C.
If the contraction proceeds while the fixed boundary condition on the bottom is maintained, the lateral side breaks near the bottom before the division of the cell, and the fixed boundary condition can not persist.
Hence we need to consider the displacement of the layer with respect to the bottom to deal with this problem correctly.

In our one-dimensional model, we break a spring when its energy exceeds a critical value.
We assume that the corresponding critical energy density is independent of both $L$ and $H$.
As mentioned above, this is equivalent to the critical stress condition in one-dimensional models.
Representing the critical energy density with the corresponding linear shrinking rate $s_b$ by $k_1s_b^2/2$, the fragmentation conditions are described by the rules
\begin{eqnarray}
e_1(x)\geq \frac{1}{2}k_1 s_b^2 \Rightarrow
\begin{minipage}{15em}
The horizontal spring is cut off; the cell is divided.
\end{minipage}
\label{crack condition 1}\\
e_2(x)\geq \frac{1}{2}k_1 s_b^2 \Rightarrow
\begin{minipage}{15em}
The vertical spring is cut off; the bottom of the layer breaks.
\end{minipage}
\label{crack condition 2}
\end{eqnarray}
Here we apply the same condition to the vertical springs in order to enforce that the lateral side breaks before the division of a cell under the fixed boundary condition.

We can easily determine the analytical solutions.
The functional variation of the energy (\ref{1d energy}) on $u(x)$ is obtained in the form
\begin{eqnarray}
\delta E=\int_{-L}^{L}dx\{-k_1 u_{xx}+k_2(u-w)\}\delta u\nonumber\\
+\left[k_1(u_x+s)\delta u\right]_{-L}^{L},
\label{1d energy minimum}
\end{eqnarray}
and we obtain both the balanced equation
\begin{mathletters}
\label{1d balance equation}
\begin{equation}
u_{xx}=q^2(u-w)
\end{equation}
 and the stress-free boundary condition
\begin{equation}
u_x+s=0\hspace{1em}\mbox{at $x=\pm L$}.
\end{equation}
\end{mathletters}

\

First we assume the fixed boundary condition without slip on the bottom: $w(x)=0$ for $|x|\leq L$.
The solution of the equations (\ref{1d balance equation}) is then
\begin{equation}
u(x)=-\frac{s}{q}\frac{\sinh{qx}}{\cosh{qL}}.
\label{deformation(no slip)}
\end{equation}

The deformation almost only appears near the lateral boundaries, because of exponential dumping.
The energy densities of horizontal $e_1(x)$ and vertical springs $e_2(x)$ are maxima at the center of a cell $x=0$ and at the lateral boundary $x=L$, respectively, and these points have the greatest possibility of breaking.
Then energy densities are calculated as
\begin{mathletters}
\label{energy density at fix b.c.}
\begin{equation}
e_1(0)=\frac{1}{2}k_1s^2\left(1-\frac{1}{\cosh{qL}}\right)^2
\label{e_1(0) at fix b.c.}
\end{equation}
and
\begin{equation}
e_2(L)=\frac{1}{2}k_1s^2\left(\tanh{qL}\right)^2.
\end{equation}
\end{mathletters}
Although they both increase with shrinking, $e_1(0)$ is always less than $e_2(L)$.

\begin{figure}[h]
%
\begin{minipage}{7cm}
\epsfxsize=7cm \epsfbox{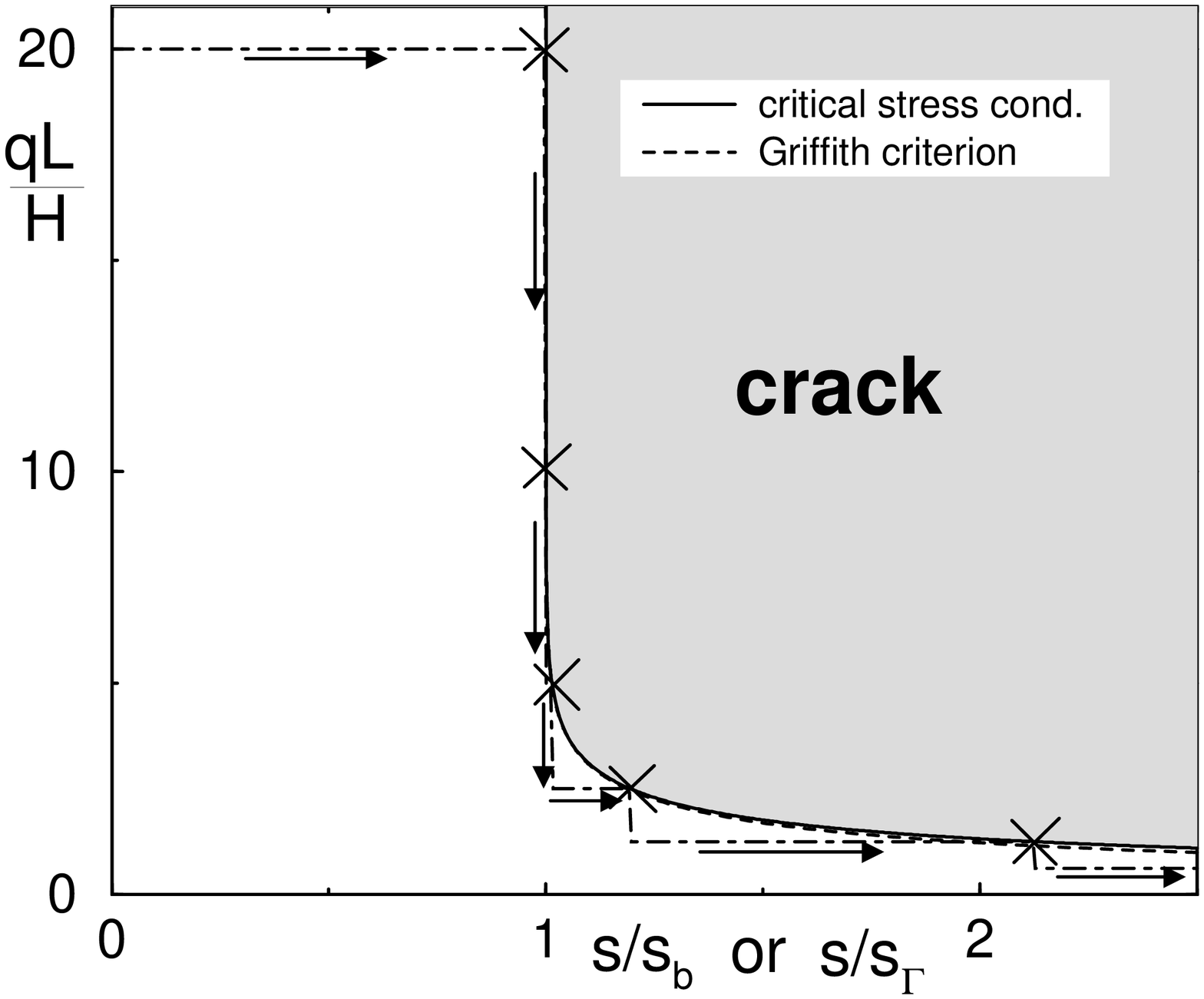}
\caption{The shrinking rate at the first breaking under the fixed boundary condition is plotted as a function of the system size, where the vertical axis represents $qL$ scaled by $H$ and the horizontal is $s$ scaled by $s_b$ or $s_{\Gamma}$.
The dotted line represents the result for the Griffith criterion.
}
\label{cracksize.fix.eps}
\end{minipage}\hspace{0.5cm}
%
\begin{minipage}{7cm}
\epsfxsize=7cm \epsfbox{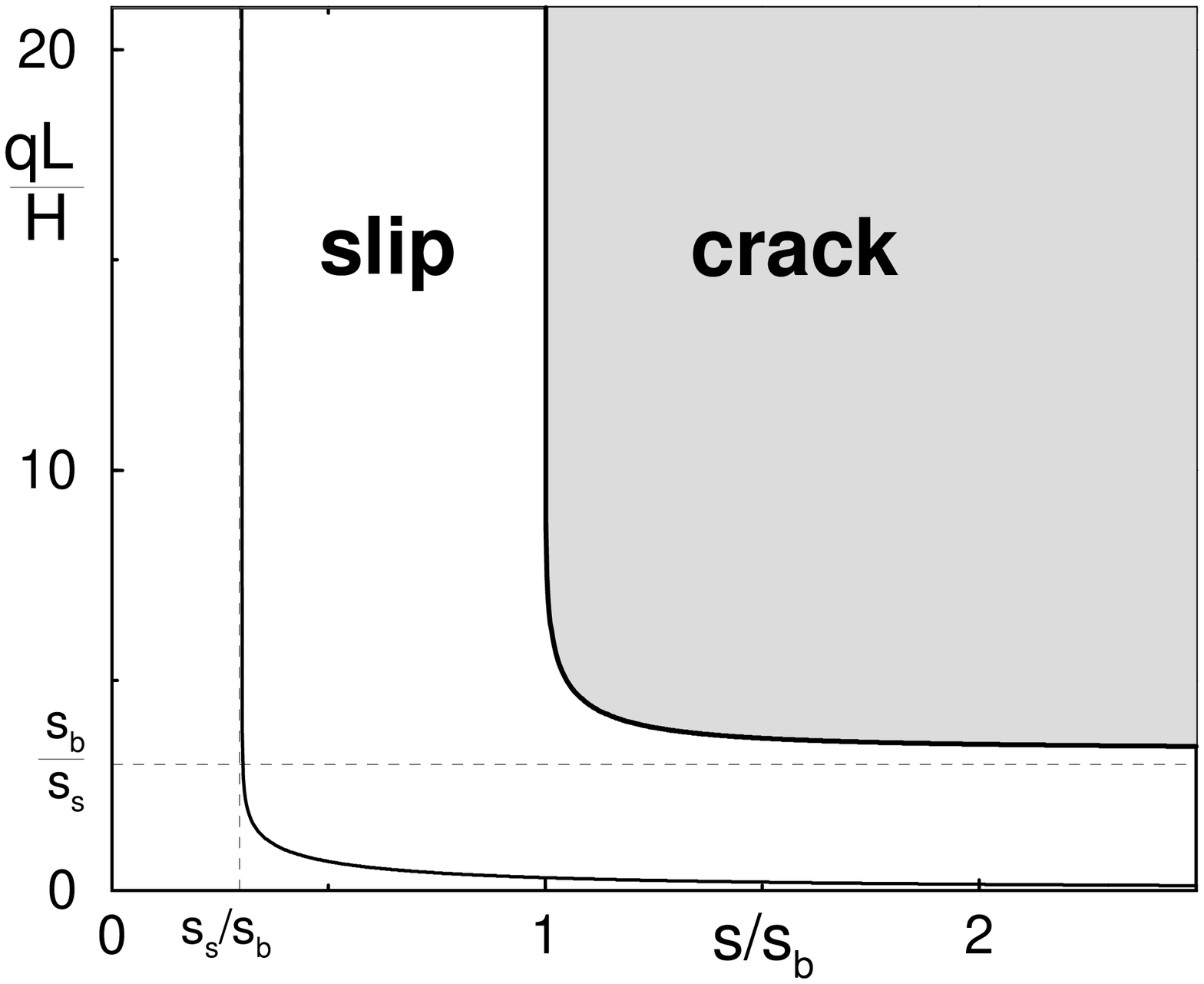}
\caption{The shrinking rates at the start of slip and at the first breaking are plotted for the system size, as in Fig. \protect\ref{cracksize.fix.eps}. }
\label{cracksize.eps}
\end{minipage}
\end{figure}

If $s_b$ is smaller than $s_s$, that is, if breaking occurs before slip, the breaking condition (\ref{crack condition 2}) for the vertical spring on the lateral side is the first to be satisfied.
To identify the effect of slip, we consider the fragmentation of a cell with the assumption that neither the slip nor the breaking of the vertical springs occurs even with the fixed boundary condition. 
For the first breaking of the horizontal springs, we apply the fragmentation condition (\ref{crack condition 1}) to (\ref{e_1(0) at fix b.c.}) and obtain the relation between the system size and the shrinking rate:
\begin{equation}
qL=\mathop{\mbox{arccosh}}{\frac{s}{s-s_b}}.
\label{crack contraction without slip}
\end{equation}
As indicated with the solid line in Fig. \ref{cracksize.fix.eps}, 
$qL$ drops rapidly at $s/s_b\simeq 1$ and then vanishes slowly as $s$ increases further.
Because each breaking divides the system into rough halves, the size of a cell decreases with shrinking.
The figure displays the typical development of the size of a cell with the stair-like function of the dot-dashed line and the arrows.
Because the system size $L$ is scaled by the thickness $H$, we see that the system is divided into a size smaller than $H$ after sufficient shrinking.

If $s_b$ is larger than $s_s$, the layer starts to slip from the lateral sides when $e_2(L)=k_1s_s^2/2$.
The shrinking rate at that time is given by
\begin{equation}
s=\frac{s_s}{\tanh{qL}}.
\label{slip contraction}
\end{equation}
We next investigate this case.

\

We suppose that the symmetrical slip from both lateral sides is directed toward the center and only consider the half region $x>0$.
The function $w(x)$ becomes finite in the slip region $x_s<x\leq L$ and remains zero elsewhere, where we introduce $x_s$ as the starting point of the slip region.
The slip displacement $w(x)\equiv w_0(x)$ is expressed by the displacement $u(x)\equiv u_0(x)$ as
\begin{equation}
w_0(x)=\left\{\begin{array}{ll}
       0                    &  0<x\leq x_s \\
       u_0(x)+\frac{s_s}{q} &  x_s<x\leq L \\
       \end{array}\right..
\label{initial slip}
\end{equation}
Equation (\ref{1d balance equation}) then take the form
\begin{mathletters}
\begin{equation}
u_{0xx}=\left\{\begin{array}{ll}
     q^2 u_0(x) &  0 < x \leq x_s \\
     -qs_s  &  x_s < x \leq L \\
     \end{array}\right.,
\end{equation}
\begin{equation}   
\hspace{2em}\mbox{b.c.:}\hspace{1em}
u_{0x}+s=0\hspace{1em}\mbox{at $x=L$ ,}
\end{equation}
and the matching conditions are
\begin{equation}
\mbox{ $w_0(x)$, $u_0(x)$ and $\frac{\partial u_0(x)}{\partial x}$ are continuous at $x=x_s$.}
\end{equation}
\end{mathletters}
We derive the solution $u_0(x)$ in each region and obtain
\begin{equation}
u_0(x)=\left\{\begin{array}{ll}
     A\sinh{qx} & 0\leq x < x_s \\
     \left[qs_s\left(L-\frac{1}{2}x\right)-s\right]x+B & x_s \leq x < L \\
     \end{array}\right..
\label{displacement for slipping}
\end{equation}
The three matching conditions give the integral constants $A$ and $B$ and yield the equation to determine $x_s$, 
\begin{equation}
q(L-x_s)=\frac{s}{s_s}-\frac{1}{\tanh{q x_s}} .
\label{x_s}
\end{equation}
At $x_s=L$, this reduces to (\ref{slip contraction}).
This form can be approximated as $L-x_s\simeq (s-s_s)/qs_s$ for $qx_s\gg1$ and as $qx_s\simeq (s/s_s-qL)^{-1}$ for $qx_s\ll1$.

We calculate the energy density $e_1(0)$ again and substitute this into the breaking condition (\ref{crack condition 1}).
This gives the equation
\begin{equation}
 \frac{s}{s_s}-\frac{1}{\sinh{q x_s}}\geq \frac{s_b}{s_s} .
\label{crack condition for s and x_s}
\end{equation}

Eliminating $x_s$ from the equations (\ref{x_s}) and (\ref{crack condition for s and x_s}), we obtain the following relation between the system size and the shrinking rate at the first breaking: 
\begin{equation}
qL=\mathop{\mbox{arcsinh}}{\left(\frac{s_s}{s-s_b}\right)}+\frac{s}{s_s}-\sqrt{1+\left(\frac{s-s_b}{s_s}\right)^2}.
\label{crack contraction}
\end{equation}
We see that $qL$ is a decreasing function of $s$.
It decreases slowly to the limiting value $s_b/s_s$ after the rapid drop in the range $s_b\leq s\lesssim s_b+s_s$.

Figure \ref{cracksize.eps} exhibits two curves of the shrinking rates at the start of slip (\ref{slip contraction}) and at the first breaking (\ref{crack contraction}), where half of the system size $L$ is represented on the vertical axis as in Fig. \ref{cracksize.fix.eps}.
After full contraction, the final size of a cell is close to the asymptotic value of the curve defined by (\ref{crack contraction}), $qL\sim s_b/s_s$.
The region without slip also becomes smaller, and its final size is given by $qx_s\simeq s_s/(s-s_b)$, where we assume $qx_s \ll 1$ in (\ref{x_s}).
With the original scale, we obtain
\begin{equation}
L \simeq \frac{H}{q} \frac{s_b}{s_s} 
\hspace{1em}\mbox{and}\hspace{1em}
x_s\simeq \frac{H}{q}\frac{s_s}{s-s_b}\hspace{2em}\mbox{for $s-s_b\gtrsim s_s$}.
\label{final cell size}
\end{equation}
Thus $L$ and $x_s$ are proportional to the thickness of a layer $H$, although there is the possibility for them to be modified through $s_s$ if the frictional force depends on $H$.

The first equation of (\ref{final cell size}) is consistent with the experimental results of Groisman and Kaplan \cite{Groisman94} for the final size of a cell after full desiccation, as mentioned in Sec. \ref{Introduction}.
The assumptions used in this analysis are also consistent with those in their qualitative explanation, where they considered the balance between the frictional force and elastic force \cite{Groisman94}.
In addition, when we peel the layer of an actual mixture after drying, we often observe a circular mark at the center of each crack cell on the bottom of the container.
Its size is approximately equal to the thickness of the layer.
We can understand these marks as the sticky region $|x|<x_s$.

\subsection{The Griffith Criterion}

Next we apply the Griffith criterion \cite{Griffith1920} to the entire system as the fragmentation condition in the place of the critical stress condition.
This was used by Sasa and Komatsu in a different model \cite{Sasa97}.

First we again assume the fixed boundary condition, where neither the slip nor the breaking of the vertical springs occurs.
The Griffith criterion introduces the creation energy of a crack surface per unit area $\Gamma$ and assumes the cracking condition that the sum of the creation energy and the elastic energy decreases due to breaking.
We write the elastic energy of a system $-L\leq x \leq L$ as $E(2L)$.
We consider the case in which the cell with size $2L$ (the system size) is divided into exact halves.
The alternative fragmentation condition to (\ref{crack condition 1}) is given by the equation
\begin{equation}
\Delta E(2L)\equiv E(2L)-2E(L)\geq\Gamma H .
\label{Griffith criterion}
\end{equation}

We calculate (\ref{1d energy}) by using (\ref{deformation(no slip)}) to obtain the elastic energy $E(2L)$ for the fixed boundary condition.
With the original scale, it is given by
\begin{equation}
E(L)=k_1 s^2 HL\left(1-\frac{H}{qL}\tanh{\frac{qL}{H}}\right).
\end{equation}

As a result, we obtain the following relation in the place of (\ref{crack contraction}) for the shrinking rate at the first breaking: 
\begin{equation}
\Delta \tilde{E}(L)=\left(\frac{s_{\Gamma}}{s}\right)^2
\hspace{1em}\mbox{and}\hspace{1em}
s_{\Gamma}\equiv \sqrt{\frac{q\Gamma}{k_1 H}} .
\label{Griffith condition on f.b.c.}
\end{equation}
Here, 
\begin{eqnarray}
\Delta \tilde{E}(L)\equiv \frac{q \Delta E(L)}{k_1 s^2H^2}=2\tanh{\frac{qL}{2H}}-\tanh{\frac{qL}{H}}\nonumber\\
=\left\{\begin{array}{ll}
 1 & qL \gg H \\
 \frac{1}{4}\left(\frac{qL}{H}\right)^3    & qL \ll H \\
 \end{array}\right..
\label{Delta E}
\end{eqnarray}
The corresponding curve is indicated with the dotted line in Fig. \ref{cracksize.fix.eps}, where the shrinking rate $s$ is scaled by $s_{\Gamma}$.
This curve agrees quite well with the solid line representing the previous results (\ref{crack contraction without slip}), so we again find that cells are divided into a size smaller than $H$ after breaking.
We however note that $s_{\Gamma}$ depends on the thickness $H$, although both $s_b$ and $s_{\Gamma}$ represent the shrinking rate at the first breaking for an infinite system.
Because the ratio of the surface energy $\Gamma$ to the elastic constant is a microscopic length for ordinary materials, $s_{\Gamma}$ is inferred to be very small.
Therefore, with the Griffith criterion, we usually expect that $s_{\Gamma}$ is smaller than $s_s$ and no slip occurs before breaking.

\

Next we show that, even if $s_{\Gamma}$ is larger than $s_s$, the Griffith criterion does not yield the proportionality relation of the final size of a cell to the thickness of the layer $H$.
Elastic energy is consumed not only by the creation of the crack surface but also by the friction due to slip on the bottom.
We again consider the breaking of the system ($-L<x<L$) into exact halves.
An alternative Griffith condition is given by 
\begin{equation}
\Delta E_s(L)\equiv E_s(2L)-2[E'_s(L)+W_s]\geq\Gamma H,
\label{Griffith criterion for slipping}
\end{equation}
where $E_s(2L)$ and $2E'_s(L)$ represent the elastic energies of the system before and after breaking, respectively, and $2W_s$ is the work performed by the frictional force due to slip. 

As the state just before breaking, we consider a cell with symmetric slip regions.
This state has been derived in (\ref{initial slip}), (\ref{displacement for slipping}) and (\ref{x_s}).
The elastic energy $E_s(2L)$ is obtained by calculating (\ref{1d energy}) in the form
\begin{equation}
E_s(2L)=\frac{k_1s_s^2}{q}\left[\frac{1}{3}q^3(L-x_s)^3+\left(\frac{s}{s_s}\right)^2q x_s-\frac{s}{s_s}\right],
\label{initial energy for slipping case}
\end{equation}
where the width of the slip region $L-x_s$ is determined as a function of $L$ and $s/s_s$ by (\ref{x_s}).

In order to estimate $W_s$ and $E'_s(L)$, we need to investigate the detailed process of fragmentation.
Here we imitate an actual quasi-static fracture in two dimensions by using a hypothetical quasi-static process in the one-dimensional model.
We introduce a traction force on crack surfaces which prevents the crack from opening and obtain the final state of this process with the stress-free boundaries by stipulating that the strength vanishes quasi-statically.
The work of the hypothetical traction force is considered to be the opposite of the creation energy of the crack.
Let us imagine the right half $0<x<L$ just after the breaking at the center $x=0$, where the traction force works at $x=0$ to the left.
Because of the relaxation of the traction, the slip region $x_s<x<L$ before the breaking vanishes immediately.
As the traction decreases, a new slip region is created in $0<x<x_r$ on the side of the crack.
If the contraction ratio $s$ is much larger than $s_s$, we may assume that $x_s$ is smaller than $x_r$ at the end of this process, because $qx_s\lesssim 1$, and the new slip region expands to $x_r \simeq L/2$.
The slip displacement $w(x)$ in the state is given by the initial condition (\ref{initial slip}) and the slip condition (\ref{slip condition}) as
\begin{equation}
w(x)=\left\{\begin{array}{ll}
       w_0(x)             &  x_r<x<L \\
       u(x)-\frac{s_s}{q} &  0<x<x_r \\
       \end{array}\right..
\label{final slip}
\end{equation}

The solution of (\ref{1d balance equation}) is 
\begin{equation}
u(x)=\left\{\begin{array}{ll}
       u_0(x)+C'_1\cosh{q(x-L)} &  x_r<x<L \\
       \frac{1}{2}qs_sx^2-sx+C'_2 &  0<x<x_r \\
       \end{array}\right.,
\label{final displacement}
\end{equation}
where the conditions of the continuity of $w(x)$, $u(x)$ and $u_x(x)$ at $x=x_r$ determine the constants $C'_1$ and $C'_2$ and produce the equation for $x_s$:
\begin{equation}
  q(L-2x_r)=2\tanh{q(L-x_r)}.
  \label{x_r}
\end{equation}
We calculate the elastic energy (\ref{1d energy}) from (\ref{final slip}) and (\ref{final displacement}) and obtain the energy at the end of this process,  
\begin{equation}
2E'_s(L)=\frac{k_1s_s^2}{q}\left\{\frac{1}{3}q^3[x_r^3+(L-x_r)^3]-qL\right\}.
\label{final energy for slipping case}
\end{equation}
Because slip occurs with a constant frictional force from the previous assumption, the work $W_s$ can be expressed by the integral of the total distance of slip,
\begin{equation}
W_s=\frac{F_s}{a}\int^{L}_{0}dx|w(x)-w_0(x)|,
\end{equation}
and it is calculated from (\ref{q and F_s}), (\ref{final slip}) and (\ref{final displacement}) as
\begin{equation}
2W_s=\frac{k_1s_s^2}{q}\left[\frac{2}{3}q^3(x_s^3-2x_r^3)+q^3Lx_r^2
                   -\left(qL-\frac{s}{s_s}\right)q^2x_s^2\right].
\label{dissipation energy for slip}		   
\end{equation}

In order to know the scaling relation of the final size of a crack cell after full desiccation, we assume $qL\gg 1$ and the limit of the full contraction: $s/s_s\rightarrow \infty$.
Because (\ref{x_s}) and (\ref{x_r}) give the approximate equations $qx_s\simeq s_s/s \ll 1$ and $x_r\simeq L/2$, respectively, (\ref{Griffith criterion for slipping}),(\ref{initial energy for slipping case}),(\ref{final energy for slipping case}) and (\ref{dissipation energy for slip}) result in the equation
\begin{equation}
\Delta E_s(L)\simeq \frac{k_1s_s^2}{6 q} (qL)^3.
\end{equation}
With the original scaling, the Griffith criterion (\ref{Griffith criterion for slipping}) gives the scaling relation of the final size of a cell for the thickness $H$, 
\begin{equation}
qL\gtrsim\sqrt[3]{6}H\left(\frac{s_{\Gamma}}{s_s}\right)^{\frac{2}{3}}\propto H^{\frac{2}{3}},
\end{equation}
where we use $s_{\Gamma}$ defined in (\ref{Griffith condition on f.b.c.}), and the condition $s_{\Gamma}\gg s_s$ is necessary from the assumption $qL\gg 1$.
Thus the Griffith criterion gives the different scaling relation because of the dependence of $s_{\Gamma}$ on $H$, although we obtained the proportionality relation (\ref{final cell size}) under the critical stress condition.

\

As a result, the critical stress condition and the Griffith condition lead to different relations between the final size of a crack cell and the thickness of a layer.
Experimental results seem to support the former.
Although the results should be discussed further, of course, they suggest  the possibility that the dissipation in the bulk can not be neglected for the fracture of a mixture of granules and water.

\section{The development of cracks in two dimensions} \label{2d model-1}

The one-dimensional model we have discussed to this point idealizes the process of cracking, treating cracks as one-dimensional structures forming one at a time parallel to one another.
In order to consider the development of cracks and their pattern formation, we must extend this model to two dimensions and include the relaxation process of the elastic field.

Although mixtures containing granular materials that are rich with water generally possess viscoelasticity, visible fluidity can not be observed at the time of cracking after the evaporation of water with desiccation.
Therefore, we assume that only the relaxation of strain contributes to the dissipation process in a linear elastic material.
For simplicity, we assume that the system has the only one characteristic relaxation time.

 \
 
In the one-dimensional model, the total energy (\ref{1d energy}) consists of terms representing a one-dimensional linear elastic material and the vertical shearing strain, the both of which are quadratic in the displacements.
Expanding the elastic material to the horizontal $xy$ plain, we naturally obtain the extended energy in two dimensions as 
\begin{mathletters}
\label{2d elastic energy}
\begin{eqnarray}
E=\int dxdy\{e_1(x,y)+e_2(x,y)\},\\
e_1(x,y)\equiv\frac{1}{2}\kappa(u_{ll}+C)^2+\mu\left(u_{ik}-\frac{1}{2}u_{ll}\delta_{ik}\right)^2
\label{2d linear elastic meterial}\\
\mbox{and}\hspace{1em}
e_2(x,y)\equiv\frac{1}{2}k_2({\bf u}-{\bf w})^2,
\end{eqnarray}
\end{mathletters}
where $u_{ij}\equiv (u_{i,j}+u_{j,i})/2$ and $u_{i,j}\equiv \partial u_i/\partial x_j$.
In analogy to the one-dimensional model, the two-dimensional vector fields ${\bf u}(x,y)$ and ${\bf w}(x,y)$ represent the displacement of a layer from the initial position and the slip displacement on a bottom, respectively.
The space coordinates $x$ and $y$ and the displacements ${\bf u}(x,y)$ and ${\bf w}(x,y)$ are again scaled by the thickness of a layer, $H$. 
The expression in (\ref{2d linear elastic meterial}) is the energy density of the two-dimensional linear elastic material with a uniform free surface-shrinking rate $C$.

The shrinking speed $\dot{C}$ can be neglected from the assumption of quasi-static contraction.
The time derivative of (\ref{2d elastic energy}) is given by 
\begin{mathletters}
\label{time derivation of energy}
\begin{eqnarray}
\dot{E}=\int dxdy\{[-\sigma_{ij,j}+k_2(u_i-w_i)]\dot{u}_i\nonumber\\
-k_2(u_i-w_i)\dot{w}_i\}+\oint dS n_j\sigma_{ij}\dot{u}_i, 
\\
\sigma_{ij} \equiv (\lambda u_{ll}+\kappa C)\delta_{ij}+2\mu u_{ij}
\hspace{1em}\mbox{and}\hspace{1em}
\kappa\equiv \lambda+\mu,
\end{eqnarray}
\end{mathletters}
where $\oint dS$ represents the integral along the boundary of the cell and ${\bf n}$ is its normal vector.

The dissipation of energy arises from the non-vanishing relative velocities of neighboring elements in a material, and then the time derivative $\dot{E}$ can also be represented as a function of them.
As is well known, $\dot{E}$ can be written in a form similar to the energy due to certain symmetries \cite{LandauElasticTheory}.
We have
\begin{mathletters}
\label{2d energy dissipation} 
\begin{eqnarray}
\dot{E}=-\int dxdy\{
\kappa' \dot{u}_{ll}^2+2\mu'\left(\dot{u}_{ik}-\frac{1}{2}\dot{u}_{ll}\delta_{ik}\right)^2\nonumber\\
+k_2'(\dot{{\bf u}}-\dot{{\bf w}})^2\} \nonumber \\
=-\int dxdy\{[-\sigma'_{ij,j}+k'_2(\dot{u}_i-\dot{w}_i)]\dot{u}_i\hspace{2em}\nonumber\\
-k'_2(\dot{u}_i-\dot{w}_i)\dot{w}_i\}-\oint dS n_j \sigma'_{ij}\dot{u}_i
\end{eqnarray}
to the second order, where 
\begin{eqnarray}
 \sigma'_{ij} \equiv \lambda' \dot{u}_{ll}\delta_{ij}+2\mu' \dot{u}_{ij}
\hspace{1em}\mbox{and}\hspace{1em}
 \kappa' \equiv \lambda'+\mu'.
\end{eqnarray}
\end{mathletters}
Although the constants $\kappa'$, $\mu'$ and  $k'_2$ are generally independent of $\kappa$, $\mu$ and $k_2$, we assume they take simple forms with one relaxation time $\tau$, writing $\sigma'_{ij}=\tau \partial \sigma_{ij}/\partial t$, or equivalently,  
\begin{equation}
\kappa'=\tau\kappa,
\hspace{1em}\mu'=\tau\mu
\hspace{1em}\mbox{and}\hspace{1em}
k'_2=\tau k_2.
\label{relaxation time}
\end{equation}

Equations (\ref{time derivation of energy}), (\ref{2d energy dissipation}) and (\ref{relaxation time}) yield the time evolution equation of ${\bf u}(x,y)$, 
\begin{equation}
\left(1+\tau\frac{\partial}{\partial t}\right)[\sigma_{ij,j}-k_2(u_i-w_i)]=0,
\label{relaxation model}
\end{equation}
and the free boundary condition, 
\begin{equation}
\left(1+\tau\frac{\partial}{\partial t}\right)\sigma_{ij}n_j=0 .
\label{2d free boundary condition}
\end{equation}
As mentioned above, the shrinking rate $C$ only appears in the free boundary condition.

Except for the effect of shrinking and slip, the above equations are essentially the same as the Kelvin model, proposed for viscoelastic solids.
Because the existence of a bottom causes a screening effect through the term $k_2 u_i$, the elastic field decays exponentially in the range of the thickness of a layer, i.e, the unit length in (\ref{relaxation model}).
Here the stress in the material is $(1+\tau \partial/\partial t)\sigma_{ij}$ by adding the dissipative force.
With the definitions
\begin{mathletters}
\label{U,W and Sigma}
\begin{equation}
U_i\equiv \left(1+\tau\frac{\partial}{\partial t}\right)u_i,\hspace{1em}
W_i\equiv \left(1+\tau\frac{\partial}{\partial t}\right)w_i
\label{U,W}
\end{equation}
and 
\begin{equation}
\Sigma_{ij}\equiv \left(1+\tau\frac{\partial}{\partial t}\right)\sigma_{ij},
\label{Sigma}
\end{equation}
\end{mathletters}
(\ref{relaxation model}) and (\ref{2d free boundary condition}) can be rewritten as 
\begin{mathletters}
\label{U equation}
\begin{eqnarray}
\Sigma_{ij,j}=k_2 (U_i-W_i),\\
\Sigma_{ij}=(\lambda U_{ll}+\kappa C)\delta_{ij}+2\mu U_{ij},
\end{eqnarray}
doing with the free boundary condition
\begin{equation}
\hspace{2em}n_j\Sigma_{ij}=0.
\end{equation}
\end{mathletters}
Therefore, $U_i$ satisfies the balanced equations of an ordinary elastic material without dissipation.

\

We expect that the state of the water bonds in a mixture can be represented by $\sigma_{ij}$, i.e. the stress excluding the dissipative force, rather than by the stress $\Sigma_{ij}$ itself because $\sigma_{ij}$ is a function of the strain $u_{ij}$.
For this reason we introduce the breaking condition by using $\sigma_{ij}$ in the following analysis.
The propagation of a crack in the Kelvin model has been investigated by many peoples \cite{Barber89,Langer92,Langer93,Ching96a,Ching96b,Kessler98}.
Although the stress field diverges at a crack tip in the continuous Kelvin model, it is possible that the divergence of $\sigma_{ij}$ is suppressed by the advance of a crack.
We calculate the elastic field around a crack tip for a straight crack which propagates stationarily in an infinite system.
Because near the tip of a propagating crack there is little slip, as the simulations in the next section indicate, we can assume ${\bf w}\simeq 0$, and then $W_i(x,y)=0$ in (\ref{U equation}).
Although our model is incomplete in the sense that the divergence of the stress can not be removed in continuous models of a linear elastic material, we expect that the following discussions are valid.

We consider a straight crack with velocity $v$ that coincides with the semi-infinite part of the $x$-axis satisfying $x<vt$ in two-dimensional plane.
The stress field satisfies the stress-free boundary conditions on the crack surface, and the displacement ${\bf u}$ vanishes as $|y| \rightarrow \infty$.
We define the moving coordinates $(\xi,y)$ as $\xi\equiv x-vt$ and assume reflection symmetry on the x-axis.
The boundary conditions are given by
\begin{eqnarray}
\left\{\begin{array}{ll}
\Sigma_{yy}=0 & \mbox{on $y=0$, $\xi<0$} \\
U_y = 0 & \mbox{on $y=0$, $\xi>0$ } \\
\Sigma_{\xi y}=0 & \mbox{on $y=0$} \\
U_i \rightarrow 0 & \mbox{for $|y| \rightarrow \infty$ } \\
\end{array}\right..
\end{eqnarray}

The stress field under the above boundary conditions can be obtained by the Wiener-Hopf method.
Fortunately, this problem reduces to the following solved problem for the mode I type of a crack.
We consider a stationary crack along the negative $x$-axis in completely linear elastic material without contraction, where we include the inertial term with the mass density $\rho$.
When a uniform pressure $\sigma^{*}$ is added on the surface of the crack from time $t=0$, the stress field $u^0_i$ is given by the equations
\begin{equation}
\sigma^0_{ij,j}=\rho\frac{\partial^2 u^0_i}{\partial t^2}
\hspace{1em}\mbox{and}\hspace{1em}
\sigma^0_{ij}\equiv \lambda u^0_{ll}\delta_{ij}+2\mu u^0_{ij}\\
\label{a solved problem}
\end{equation}
and the boundary conditions
\begin{equation}
\left\{\begin{array}{ll}
\sigma^0_{yy}=-\sigma^*\Theta(t) & \mbox{on $y=0$, $x<0$} \\
u^0_y = 0 & \mbox{on $y=0$, $x>0$ } \\
\sigma^0_{xy}=0 & \mbox{on $y=0$} \\
u^0_i \rightarrow 0 & \mbox{for $|y| \rightarrow \infty$ } \\
\end{array}\right..
\label{b.c. of a solved problem}
\end{equation}
These become identical to the previous equations when we apply the Laplace transformations on time, 
\begin{mathletters}
\begin{equation}
U_i(x,y,\eta)=\int_{0}^{\infty}dt u^0_i(x,y,t)e^{-\eta t}
\end{equation}
and
\begin{equation}
\Sigma^0_{ij}(x,y,\eta)=\int_{0}^{\infty}dt \sigma^0_{ij}(x,y,t)fe^{-\eta t}, 
\end{equation}
\end{mathletters}
and make the replacements $\eta^2\rho=k_2$,\ $\sigma^*=\kappa C\eta$,\ $\Sigma_{ij}=\Sigma^0_{ij}+\kappa C\delta_{ij}$ and  $x\rightarrow \xi$.
Here $\eta$ can be taken equal to 1 because the correspondence holds for any $\eta$.

Using the analytical solutions \cite{Freund90} of (\ref{a solved problem}) and (\ref{b.c. of a solved problem}), we shall consider the stress in front of the crack, i.e. $\Sigma_{yy}(\xi,0)$ where $\xi>0$.
The above replacements give the solution of our problem as,
\begin{equation}
\Sigma_{yy}(\xi,0)=\kappa C\left\{-\frac{1}{\pi}\int_{\alpha-0i}^{\infty-0i}d\zeta\ Im \hat{\Sigma}^0(-\zeta) \ e^{-\zeta \xi}+1 \right\},
\label{Sigma yy}
\end{equation}
\begin{equation}
\alpha\equiv \sqrt{\frac{k_2}{\lambda+2\mu}}
\hspace{1em}\mbox{and}\hspace{1em}
\hat{\Sigma}^0(\zeta)\equiv \frac{1}{\zeta}\left(\frac{F_+(0)}{F_+(\zeta)}-1\right)
\end{equation}
where $F_+(\zeta)$ is an analytical function in the region $Re \zeta> -\alpha$ such that 
\begin{equation}
F_+(\zeta)\rightarrow \zeta^{-\frac{1}{2}}\hspace{1em}\mbox{for $|\zeta| \rightarrow \infty$}
\end{equation}
and
\begin{equation}
F_+(0)=\sqrt{\frac{2\mu(\lambda+\mu)}{\alpha(\lambda+2\mu)^2}}.
\end{equation}
Because $Im \hat{\Sigma}^0(-\zeta)$ approaches $-F_+(0)/\sqrt{\zeta}$ as $|\zeta|$ increases, it can be approximated in the region $\alpha\xi \ll 1$ around the crack tip by the equation
\begin{eqnarray}
\Sigma_{yy}(\xi,0)\simeq \kappa C\left (\frac{F_+(0)}{\pi}\int_{0}^{\infty} d\zeta \zeta^{-\frac{1}{2}}e^{-\zeta \xi}+1\right)\nonumber\\
=\kappa C\left (\frac{F_+(0)}{\sqrt{\pi\xi}}+1\right).
\end{eqnarray}
We see that the stress $\Sigma_{yy}$ diverges in inverse proportion to the square root of the distance near the tip. 

Solving (\ref{Sigma}) in the moving system $\xi=x-vt$, we get 
\begin{equation}
 \sigma_{yy}(\xi,0)=\frac{1}{\tau v}\int_{\xi}^{\infty}d\xi'\Sigma_{yy}(\xi',0)e^{\frac{\xi-\xi'}{\tau v}},
\hspace{1em}\mbox{for $\xi>0$.}
\end{equation}
and the following equation is obtained by substituting (\ref{Sigma yy}) into this equation: 
\begin{equation}
 \sigma_{yy}(\xi,0)=\kappa C\left\{-\frac{1}{\pi}\int_{\alpha-0i}^{\infty-0i}d\zeta\ \frac{Im \hat{\Sigma}^0(-\zeta)}{\tau v \zeta+1}e^{-\zeta \xi}+1 \right\}.
 \label{sigam yy}
\end{equation}

We first examine the behavior of $\sigma_{yy}$ in the region $\alpha\xi\gg 1$ far away from the tip.
Because the integral in (\ref{sigam yy}) can be approximated as
\begin{eqnarray}
  \int_{\alpha-0i}^{\infty-0i}d\zeta\ \frac{Im \hat{\Sigma}^0(-\zeta)}{\tau v \zeta+1}e^{-\zeta \xi}\hspace{7em}\nonumber\\
=\left[\int_{0}^{\infty}d\zeta\ \frac{Im \hat{\Sigma}^0(-\zeta-\alpha)}{\tau v (\zeta+\alpha)+1}e^{-\zeta \xi}\right]e^{-\alpha\xi}\nonumber\\
\simeq
\frac{Im \hat{\Sigma}^0(-\alpha)}{\tau v \alpha+1}\frac{e^{-\alpha\xi}}{\xi},
\hspace{7.5em}
\end{eqnarray}
$\sigma_{yy}$ decays exponentially to the uniform tension as
\begin{mathletters}
\begin{equation}
 \sigma_{yy}(\xi,0)\simeq \kappa C\left(\frac{D}{\xi}e^{-\alpha\xi}+1 \right)\hspace{1em}
 \mbox{for $\alpha\xi \gg 1$},
\end{equation}
where
\begin{equation}
 D\equiv -\frac{Im \hat{\Sigma}^0(-\alpha)}{\pi(\tau v \alpha+1)} .
\end{equation}
\end{mathletters}
The characteristic length of the decay is $H/\alpha$ with the original scale, which is of the same order as the thickness $H$ for an ordinary elastic material.

In the region $\alpha\xi\ll 1$ near the tip, we approximate the integral with the asymptotic form of $Im \hat{\Sigma}^0(-\zeta)$ for large $\zeta$ as 
\begin{mathletters}
\begin{eqnarray}
-\frac{1}{\pi}\int_{\alpha-0i}^{\infty-0i}d\zeta\ \frac{Im \hat{\Sigma}^0(-\zeta)}{\tau v \zeta+1}e^{-\zeta \xi}\hspace{4em}\nonumber\\
\simeq 
\int_{0}^{\infty}d\zeta\ \frac{F_+(0)}{\pi(\tau v \zeta+1)\zeta^{\frac{1}{2}}}e^{-\zeta \xi}\nonumber\\
=\frac{2F_+(0)}{\sqrt{\pi \tau v}}e^{\frac{\xi}{\tau v}} {\rm erfc}\left(\sqrt{\frac{\xi}{\tau v}}\right),
\end{eqnarray}
\begin{eqnarray}
 {\rm erfc}(x)\equiv \int_x^{\infty}dt e^{-t^2}
 \simeq \frac{1}{2x}e^{-x^2}\hspace{1em}\mbox{for $x\rightarrow \infty$},\\
\mbox{and}\hspace{1em}
 {\rm erfc}(0)=\frac{\sqrt{\pi}}{2},
\end{eqnarray}
\end{mathletters}
and then we obtain the equations
\begin{equation}
 \sigma_{yy}(\xi,0)\simeq \left\{\begin{array}{ll}
 \itemsep=10mm
 \kappa C\left(\frac{F_+(0)}{\sqrt{\pi \xi}}+1 \right) & \xi \gg \tau v \\
 \kappa C\left(\frac{F_+(0)}{\sqrt{\tau v}}+1 \right)  & \xi \ll \tau v 
 \end{array}\right. \hspace{1em}\mbox{for $\alpha\xi \ll 1$.}
\label{sigma yy 0} 
\end{equation}
With the original scale again, $\sigma_{ij}$ is almost proportional to $\sqrt{H/\xi}$ in the middle region from $\tau v$ to $H/\alpha$ around the crack tip, as is the case for the stress $\Sigma_{ij}$.
However, $\sigma_{ij}$ is bounded at the tip by the proportional value to $\sqrt{H/\tau v}$.

\

Thus the stress excluding the dissipative force, i.e. $\sigma_{ij}$, is kept from diverging by the movement of the crack tip.
Therefore we can introduce a critical value $\sigma_{th}$ as a material parameter again and assume the breaking condition at the tip 
\begin{equation}
\lim_{\xi\rightarrow 0+}{\sigma_{yy}(\xi,0)}\geq \sigma_{th}\equiv \kappa C_{th},
\label{2d crack condition at a crack tip}
\end{equation}
where the constant $C_{th}$ represents the shrinking rate corresponding to the critical value.
For a stationary propagating crack, we find the velocity $v$ by substituting $\sigma_{yy}(\xi,0)$ into the above equation in (\ref{sigma yy 0}) as 
\begin{equation}
v=F_+(0)^2\left(\frac{C}{C_{th}-C}\right)^2\frac{H}{\tau} .
\label{crack velocity}
\end{equation}
Although this equation is not valid near the sound velocity because we have neglected inertia, we expect that a material cracks at a shrinking rate $C$ below $C_{th}$ due to inhomogeneity.
Thus (\ref{crack velocity}) explains why the propagation velocity observed in experiments is very small compared to the sound velocity.
In addition, it suggests the proportionality relation between the thickness and the propagating speed, which should be experimentally observable.

\

Next we note the validity of (\ref{crack velocity}) for very slow speeds.
Although the divergence of $\sigma_{yy}$ is suppressed by the advance of a crack, as we see in (\ref{sigma yy 0}), the size of the screening region is approximately $\tau v$.
Because particles of size $R= 0.01 \sim 1 mm$ are used in experiments, the above breaking condition (\ref{2d crack condition at a crack tip}) is available for velocities $v \gg R/\tau$, where the continuous approximation is valid.
For very slow velocities, $\tau v \ll R$, for example, it may be possible for the defects in a material to arrest the growing of cracks.

The speed of a real crack measures $v \lesssim 2mm/min$ for the thickness of a layer of coffee powder, $H\simeq 6mm$, as observed in experiments \cite{Groisman94}.
Although we have not yet specified the origin of the relaxation, we attempt to estimate the relaxation time $\tau$ arising from the viscosity of the water in the bonds among particles.
Supposing that the diameter of a particle of the coffee powder is about $R\sim 0.5mm$ and the viscosity of the water is $\nu\sim 1 mm^2/s$, we obtain $\tau \sim R^2/\nu \sim 0.25s$ and $\tau v/R \sim 0.02<1$.
This rough estimate suggests the possibility that the above continuous approximation is imperfect in the case of a relatively thin layer.

Groisman and Kaplan reported the interesting experimental results that the propagation speed exhibits a wide dispersion even among cracks growing at the same time, although the speed of each individually is almost constant on time.
It is a future problem to understand the relation of the dispersion to the inhomogeneity of materials.

\section{The patterns of cracks} \label{2d model-2}

Cracks appear one after another with shrinking, and spread over the system to create a two-dimensional pattern.
In this section, we report on a study of the formation of crack patterns using numerical simulations of the two-dimensional model introduced in the previous section.
We make the natural extension of both the breaking condition and the slip condition employed in the one-dimensional model by using the energy densities defined in the microscopic cells of the lattice in the simulations.
In the two-dimensional model, the former is simpler than the critical stress condition because it neglects the direction of both the stress and the microscopic crack surfaces.
In addition, the slip condition implicitly assumes a sufficiently short period of slip in comparison to the relaxation time of the elastic field because of the balance between the frictional force and elastic force.
We report the results of our simulations after describing the discrete method and these extended conditions.

\

As is the case with most fingering patterns, the growing of cracks is influenced strongly by the anisotropy of the system.
We used random lattices \cite{Friedberg84,Moukarzel92} in our simulations to consider uniform and isotropic systems in a statistical sense.

Many fracture models employ a network of springs or elastic beams to model an elastic material \cite{Louis87,Fernandez88,Herrmann89,Furukawa93,Hayakawa94a,Hayakawa94b,Hornig96,Leung97,Astrom98,Galdarelli98}.
However, it is generally difficult to calculate the elastic constants for an elastic material modeled by such a lattices.
Therefore, instead of such networks, we consider each triangular cell in a random lattice as a tile of the elastic material with uniform deformation.
A fracture is realized by removing any cell whose energy density exceeds a critical value, as we explain below.

We construct a two-dimensional model as is illustrated in Fig. \ref{lattice.eps}.
Each site of the lattice is connected to an element on the bottom with a vertical spring similar as that used in the one-dimensional model.
Figure \ref{Voronoi.eps} displays a part of the random lattice which represents a horizontal elastic plane.
The random lattice is composed of random points to form a Voronoi division of them.
We number the sites and the triangular cells in the lattice and express the area of the Voronoi cell around the $n$th site as $V_n\ (n=1,2,3,...,N)$ and the area of the $m$th triangle as $T_m\ (m=1,2,3,...,N_T)$.

The displacement of the $n$th site ${\bf u}^{(n)}(t)$ is obtained from the ${\bf U}^{(n)}(t)$ by the definition (\ref{U,W}). 
A simple Euler method gives the following equation with discrete time $\Delta t$:
\begin{equation}
{\bf u}^{(n)}(t+\Delta t)={\bf u}^{(n)}(t)+\frac{\Delta t}{\tau}({\bf U}^{(n)}(t)-{\bf u}^{(n)}(t)).
\label{U->u}
\end{equation}

Because ${\bf U}(t)$ satisfies the balanced equation of an ordinary elastic material at any time, it minimizes the energy $\tilde{E}$ defined by (\ref{2d elastic energy}) through the replacements ${\bf u}\rightarrow {\bf U}$ and ${\bf w}\rightarrow {\bf W}$.
$\tilde{E}$ consists of the energies of both the vertical springs and the horizontal elastic plain.
We calculate the latter by summation of the energies of the triangular cells in the random lattice, where the $m$th triangular cell is assumed to consist of a linear elastic material with uniform strain tensor $U^{(m)}_{ij}$.
Thus we obtain the equations
\begin{mathletters}
\label{2d discrete elastic energy}
\begin{eqnarray}
\tilde{E}=\mathop{{\sum}'}_{m=1}^{N_{T}}T_m\tilde{e}^{(m)}_{1}+\sum_{n=1}^{N}V_n\tilde{e}^{(n)}_{2},
\\
\tilde{e}^{(m)}_1\equiv\frac{1}{2}\kappa(U^{(m)}_{ll}+C)^2+\mu\left(U^{(m)}_{jk}-\frac{1}{2}U^{(m)}_{ll}\delta_{jk}\right)^2
\label{tilde e1}
\\
\mbox{and}\hspace{1em}
\tilde{e}^{(n)}_2\equiv\frac{1}{2}k_2({\bf U}^{(n)}-{\bf W}^{(n)})^2,
\label{tilde e2}
\end{eqnarray}
\end{mathletters}
where ${\bf U}^{(n)}$ and ${\bf W}^{(n)}$ represent the displacement of the $n$th site and the slip displacement of the $n$th element along the bottom, respectively.
Although the rule of repeated indices is applied to $j,k$ and $l$, as usual, the summations over $m$ and $n$ are expressed by the symbol $\sum$, where $\mathop{{\sum}'}_{m}$ represents a summation that excludes broken triangle cells.

The quantity $\tilde{e}^{(m)}_1$ is the elastic energy of the $m$th triangle cell which is calculated from $U^{(m)}_{ij}$.
For the following explanation, we express the vertices of the $m$th triangle as $n=1,2,3$ and their initial equilibrium positions as ${\bf x}^{(n)}\equiv (x^{(n)},y^{(n)})$, as is shown in Fig. \ref{Voronoi.eps}.
Assuming uniform deformation in the triangle, the strain tensor $U^{(m)}_{jk}\equiv \frac{1}{2}(U^{(m)}_{j,k}+U^{(m)}_{k,j})$ is given by the displacements of the vertices ${\bf U}^{(n)}$ as the equations
\begin{eqnarray}
U^{(m)}_{xx}=\frac{1}{T}\epsilon_{ijk}y^{(ij)}U_x^{(k)},\\
U^{(m)}_{yy}=-\frac{1}{T}\epsilon_{ijk}x^{(ij)}U_y^{(k)},\\
U^{(m)}_{xy}=\frac{1}{2T}\epsilon_{ijk}[y^{(ij)}U_y^{(k)}-x^{(ij)}U_x^{(k)}],
\end{eqnarray}
where
\begin{equation}
T\equiv \epsilon_{ijk}x^{(i)}y^{(jk)},\hspace{1em}
T_{m}=\frac{1}{2}|T|,\hspace{1em}
{\bf x}^{(ij)}\equiv {\bf x}^{(i)}-{\bf x}^{(j)},
\end{equation}
and $\epsilon_{ijk}$ is Eddington's $\epsilon$.
These equations are easily obtained from the first order Taylor expansions of ${\bf U}^{(n)}={\bf U}({\bf x}^{(n)},t)$ in the triangle.
Thus we can calculate the energy (\ref{2d discrete elastic energy}) on the random lattice and obtain ${\bf U}^{(n)}$ from its minimum.

\

A fracture is represented by the removal of triangle cells, not by the breaking of bonds, in this model.
This gives a direct extension from the one-dimensional model, although it neglects the microscopic direction of the stress and the cracking in a triangle cell.
We calculate the elastic energy density of the $m$th triangular cell $e^{(m)}$ from the true displacements ${\bf u}^{(n)}$, and assume a critical value for the breaking, which is represented by the corresponding shrinking rate $C_b$ as $\kappa C_b^2/2$.
Thus the breaking condition is given by
\begin{equation}
e^{(m)}_1 \geq \frac{1}{2}\kappa C_b^2 \Rightarrow
\begin{minipage}{12em}
The $m$th triangle cell is removed,
\end{minipage}
\label{2d crack condition}
\end{equation}
where
\begin{equation}
e^{(m)}_1\equiv\frac{1}{2}\kappa(u^{(m)}_{ll}+C)^2+\mu\left(u^{(m)}_{jk}-\frac{1}{2}u^{(m)}_{ll}\delta_{jk}\right)^2, 
\label{2d discrete energy density 1}
\end{equation}
and $u^{(m)}_{jk}$ is calculated from ${\bf u}^{(n)}$ using the method explained above for $U^{(m)}_{jk}$.

The slip condition for the elements on the bottom is also similar to that in the one-dimensional model.
We introduce the maximum frictional force as a constant and assume the balance of the fictional force against the sum of the elastic force and the dissipative force, i.e., $|k_2\left(1+\tau \partial/\partial t\right)({\bf u}-{\bf w})|=k_2|{\bf U}-{\bf W}|$.
Therefore, the slip condition for the $n$th element can be written by using $\tilde{e}_2^{(n)}$ as 
\begin{equation}
\tilde{e}^{(n)}_2\geq \frac{1}{2}\kappa C_s^2 \Rightarrow
\begin{minipage}{12em}
${\bf W}^{(n)}$ is moved along the force to the position where $\tilde{e}^{(n)}_2=\frac{1}{2}\kappa C_s^2$.
\end{minipage}
\label{2d slip condition}
\end{equation}
The slip displacement ${\bf w}^{(n)}$ is calculated from ${\bf W}^{(n)}$ by the definition (\ref{U,W}) as
\begin{equation}
{\bf w}^{(n)}(t+\Delta t)={\bf w}^{(n)}(t)+\frac{\Delta t}{\tau}({\bf W}^{(n)}(t)-{\bf w}^{(n)}(t)).
\label{W->w}
\end{equation}

\

We carried out the numerical simulations of the model by increasing the shrinking rate $C$ in proportion to time $t$ with a constant rate $\dot{C}$.
We repeated the following procedures at each time step: 
\begin{enumerate}
\itemsep=0mm
\item The contraction rate increases as $C(t+\Delta t)=C(t)+\Delta t \dot{C}$.
\item The displacement ${\bf u}$ is calculated from ${\bf U}$ which is the function minimizing the energy (\ref{2d discrete elastic energy}).
\item The slip ${\bf w}$ is calculated from ${\bf W}$, which is given by the conditions (\ref{2d slip condition}) at the all sites.
\item If some triangular cell satisfy the breaking condition (\ref{2d crack condition}), it is removed.
Its energy is not included in subsequent calculations.
\end{enumerate}
If more than one cell satisfies the breaking condition at step 4, we repeat the step 1-3 using a smaller time step.

In the above calculations, we can take the parameters $\kappa$, $k_2$, $C_b$ and $\dot{C}$ to be 1 with loss of generality by scaling space, time, energy and the shrinking rate as follows, 
\begin{eqnarray}
{\bf x}\rightarrow \sqrt{\frac{\kappa}{k_2}}{\bf x},\hspace{1em}
{\bf u}\rightarrow C_b \sqrt{\frac{\kappa}{k_2}}{\bf u},\hspace{1em}
{\bf w}\rightarrow C_b \sqrt{\frac{\kappa}{k_2}}{\bf w},\nonumber\\
E\rightarrow \frac{1}{2}\kappa C_b^2 E,\hspace{1em}
t\rightarrow \frac{C_b}{\dot{C}}t,\hspace{1em}
\mbox{and}\hspace{1em}C=C_b \hat{C}.
\label{scaling}
\end{eqnarray}
Here we write the scaled shrinking rate as $\hat{C}$.
As a result, three independent parameters remain explicitly in the equations: 
\begin{equation}
   \hat{\mu}\equiv\frac{\mu}{\kappa},\hspace{1em}
   \hat{\tau}\equiv\frac{\dot{C}}{C_b}\tau ,\hspace{1em}
   \mbox{and}\hspace{1em}
   \hat{C}_s\equiv\frac{C_s}{C_b}.
\end{equation}

Although the properties of the lattice influence the results, we used an identical lattice in all of our simulations, except the last in which we consider the effect of a random lattice.
To prepare the sites in the random lattice, we arranged points in a triangular lattice with mesh size $0.01$ inside a square region $1 \times 1$ and shifted the $x,y$ coordinates of each point by adding uniform random numbers within the range $\pm 0.005$.
Because their distribution is almost uniform and random, we connect them by Voronoi division to make the network of the random lattice.
Then the square region is extended to the size $\hat{L}\times\hat{L}$, which is related to the original system size $L$ as
\begin{equation}
   \hat{L}\equiv \sqrt{\frac{k_2}{\kappa}}\frac{L}{H}.
\end{equation}

We use the conjugate gradient method \cite{NumericalRecipes} with a tolerance $10^{-6}$ to find the minimum points of functions on free boundary conditions.
The time step $\Delta t$ is changed automatically in the range to an upper bound $10^{-4} - 10^{-3}$.
In our simulations, at most one cell is removed in any given timestep.
In contrast, we note that in the model without a relaxation process, a crack propagates instantaneously at a fixed shrinking rate, because the breaking of a cell necessarily changes the balanced state of the elastic field and often causes a chain consisting of the breaking of many cells occuring simultaneously.

\

\begin{figure}[t]
%
\begin{minipage}{7cm}
\epsfxsize=6cm \epsfbox{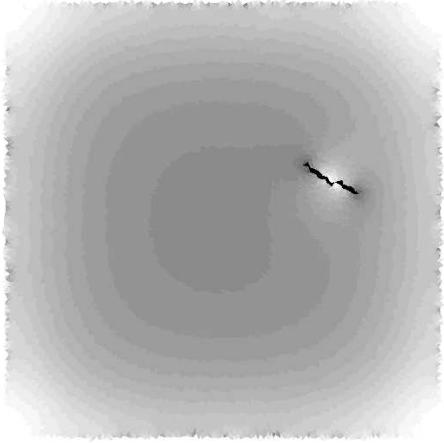}

(a)\ $\hat{C}=0.86$
\end{minipage}\hspace{1cm}
\begin{minipage}{7cm}
\epsfxsize=6cm \epsfbox{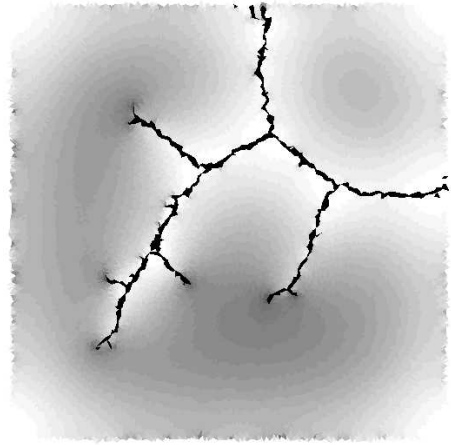}

(b)\ $\hat{C}=0.97$
\end{minipage}

\vspace{0.5mm}
\begin{minipage}{7cm}
\epsfxsize=6cm \epsfbox{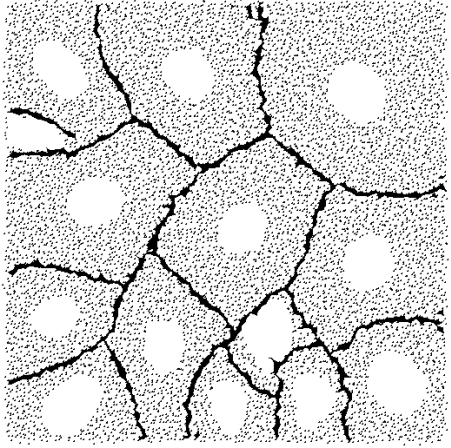}

(c) $\hat{C}=1.46$
\end{minipage}\hspace{1cm}
\begin{minipage}{7cm}
\epsfxsize=6cm \epsfbox{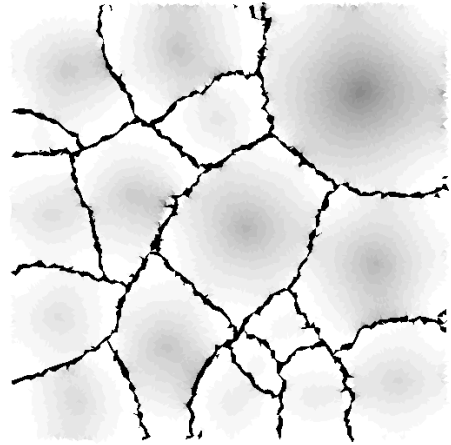}

(d) $\hat{C}=10.00$
\end{minipage}

\caption{
The time series of a crack pattern for $\hat{L}=10\protect\sqrt{2}$, $\hat{\mu}=1$, $\hat{\tau}=0.01$ and $\hat{C}_s=0.5$.
The black area represents cracks.
The gray scale in (a),(b) and (d) indicates the energy densities of the triangular cells of a lattice, and the dots in (c) indicate slipping elements.
}
\label{m10s51g12L2+1.ps}
\end{figure}

We show the typical development of a crack pattern in our model in Fig. \ref{m10s51g12L2+1.ps}, where the parameters are $\hat{L}=10\sqrt{2}$, $\hat{\mu}=1$, $\hat{\hat{\tau}}=0.01$ and $\hat{C}_s=0.5$, and the black area indicates the removed triangular cells.
The gray scale represents the energy densities (\ref{2d discrete energy density 1}) in Figs. \ref{m10s51g12L2+1.ps} (a),(b) and (d), and each dot in Fig. \ref{m10s51g12L2+1.ps} (c) is a slipping element under the condition (\ref{2d slip condition}).
$\hat{C}=1$ (i.e. $C=C_b$) corresponds to the shrinking rate at the first breaking for an infinite system.
The first breaking in the simulations occurs slightly below $\hat{C}=1$ for most parameters because of the randomness in the lattice.
At that time, no slip has yet begun in the most of the system, except near the boundaries.

After this, some crack tips grow simultaneously in the whole system, and the crack pattern almost becomes complete at the shrinking rate near $\hat{C}=1$.
Here we see the white circular marks around the center of the crack cells, as shown in Fig. \ref{m10s51g12L2+1.ps}-(c).
These represent the sticky regions without slip. 
Similar marks can be observed on the bottom of a container in actual experiments, as we mentioned in Sec. \ref{1d model-2}.

\begin{figure}[t]
%
\begin{minipage}{7cm}
\epsfxsize=6cm \epsfbox{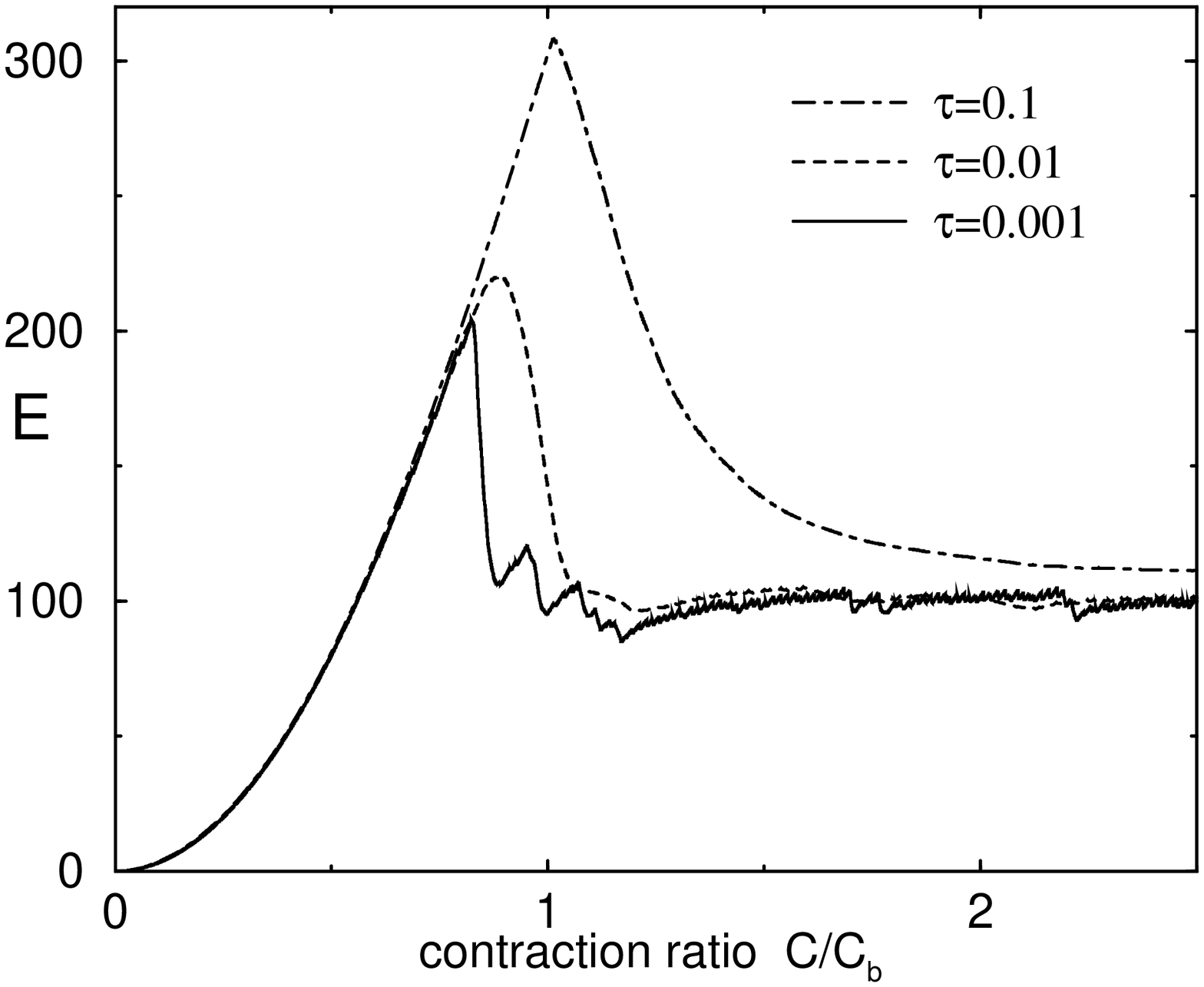}
\caption{The change of the total energy with contraction for $\hat{L}=10\protect\sqrt{2}$, $\hat{\mu}=1$, $\hat{C}_s=0.5$ and $\hat{\tau}=0.001$ for the solid line, $\hat{\tau}=0.01$ for the dotted line, and $\hat{\tau}=0.1$ for the dot-dashed line.}
\label{m10s51L2+1.energy.eps}
\end{minipage}\hspace{1cm}
%
\begin{minipage}{7cm}
\epsfxsize=6cm \epsfbox{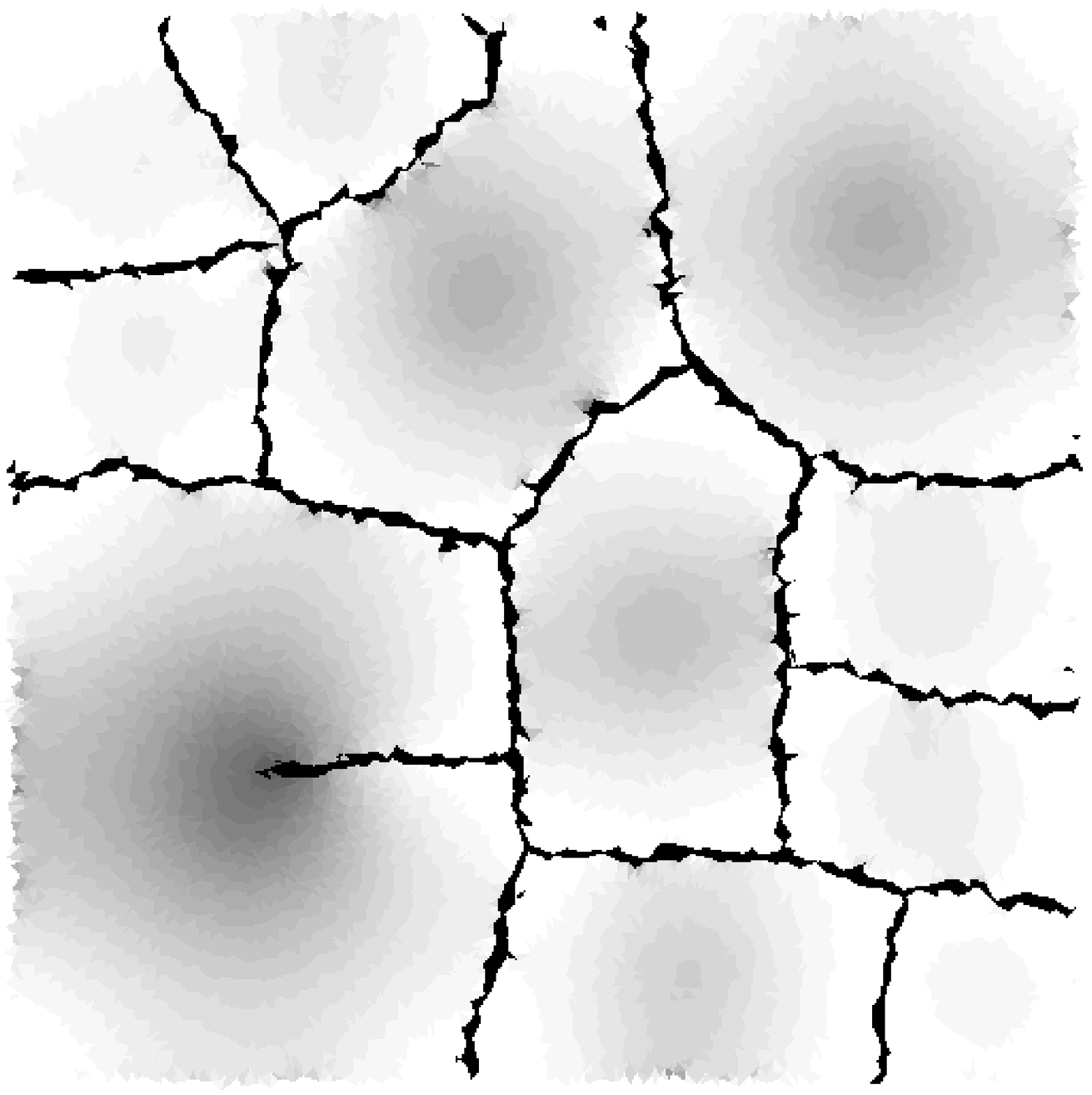}
\caption{The time development of cracks with fast relaxation.
$\hat{L}=10\protect\sqrt{2}$, $\hat{\mu}=1$, $\hat{C}_s=0.5$, $\hat{\tau}=0.001$ and $\hat{C}=1.54$.}
\label{m10s51g13L2+1.ps}
\end{minipage}

\vspace{0.5cm}
%
\begin{minipage}{7cm}
\epsfxsize=6cm \epsfbox{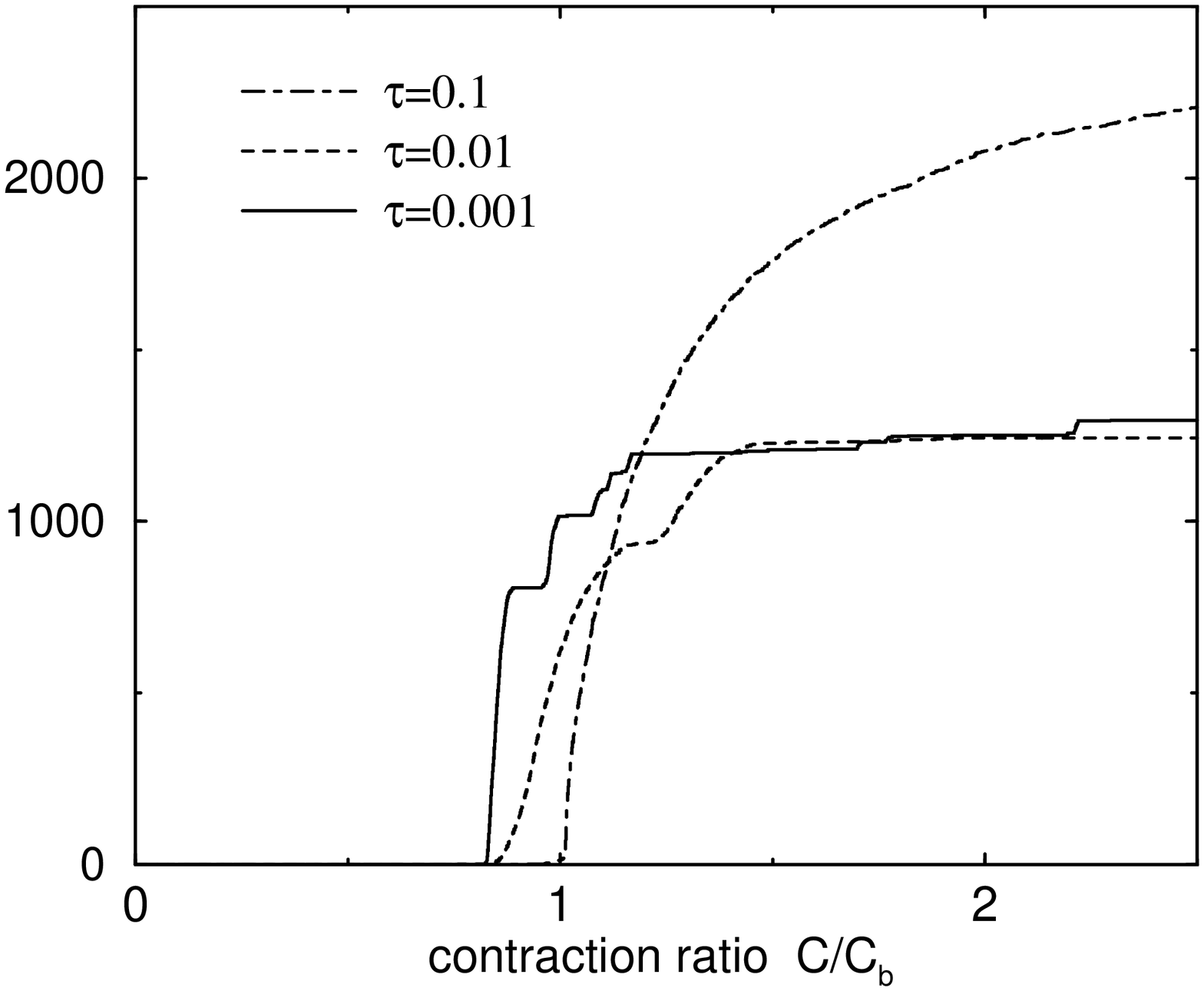}
\caption{The change of the total number of broken triangular cells for the simulations of Fig. \protect\ref{m10s51L2+1.energy.eps}.}
\label{m10s51L2+1.cellnum.eps}
\end{minipage}\hspace{1cm}
%
\begin{minipage}{7cm}
\epsfxsize=6cm \epsfbox{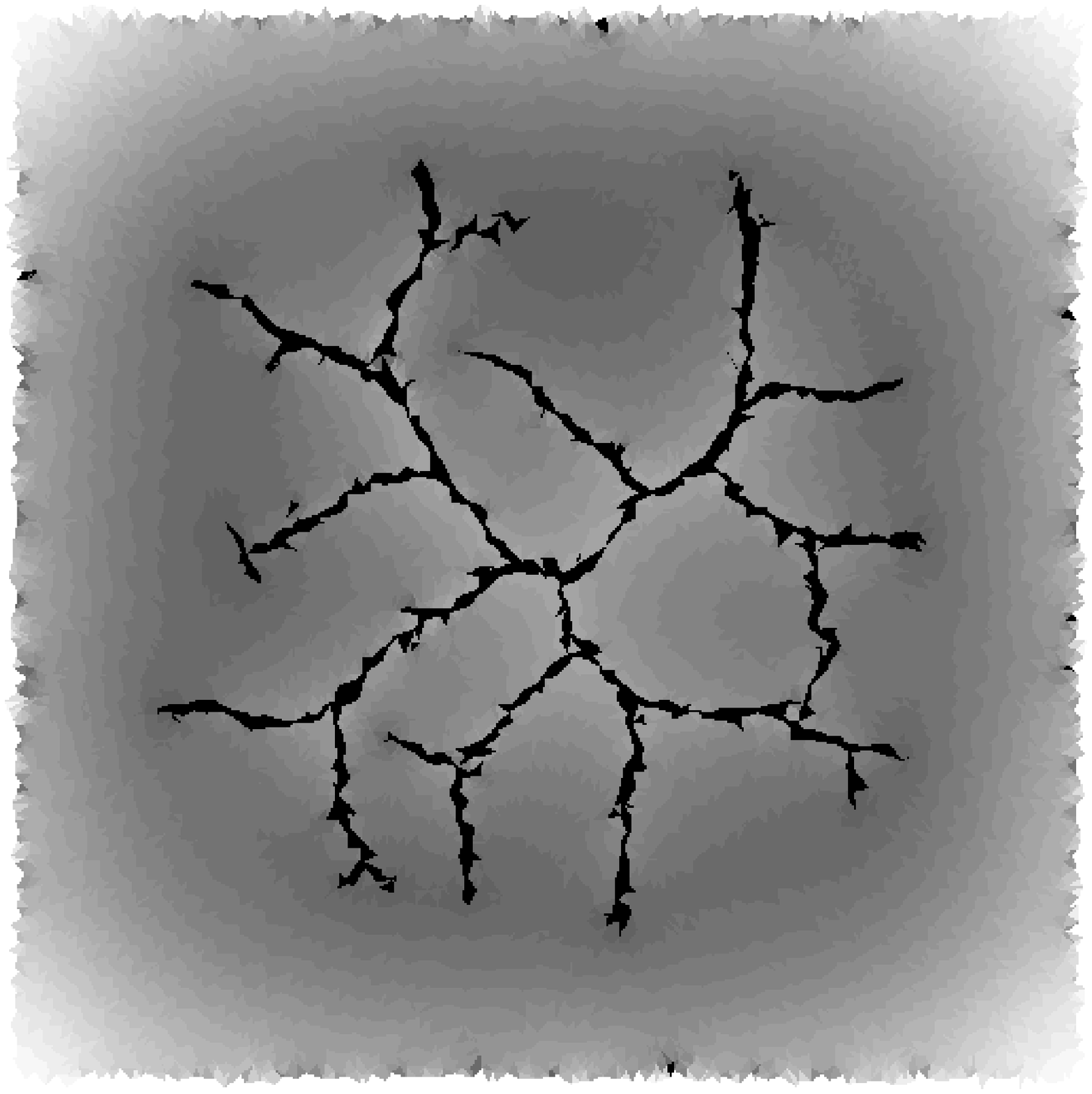}
\caption{The time development of cracks with slow relaxation. $\hat{L}=10\protect\sqrt{2}$, $\hat{\mu}=1$, $\hat{C}_s=0.5$, $\hat{\tau}=0.2$ and $\hat{C}=1.09$.}
\label{m10s51g50L2+1.ps}
\end{minipage}
\end{figure}

Figure \ref{m10s51L2+1.energy.eps} graphs the development of the total energy (\ref{2d elastic energy}) with shrinking for the three cases $\hat{\tau}=0.1$, $0.01$ and $0.001$.
The energy increases with contraction.
However, it is released and dissipates due to successive breaking and becomes almost constant with increasing shrinking rate.
Our simulations were carried out until $\hat{C}=10$.
The crack patterns changed little when $\hat{C}$ becomes large, while the circular marks shrank gradually.

\

For fast relaxation, new cracks grow from the lateral side of another crack almost perpendicularly.
Figure \ref{m10s51g13L2+1.ps} displays a snapshot of a crack pattern for the very small relaxation time $\hat{\tau}=0.001$.
As $\tau$ becomes smaller, the cracks tend to propagate faster and grow one at a time, in agreement with the assumption in the one-dimensional model.
In Fig. \ref{m10s51L2+1.cellnum.eps}, we compare how the total number of broken triangular cells increases with shrinking for $\hat{\tau}=0.1$, $0.01$ and $0.001$.
The number of broken cells is almost proportional to the total length of cracks.
It increases like a step function for $\hat{\tau}=0.001$.
This indicates that the cracks are formed one by one.

For slow relaxation, in contrast, the growing cracks from fingering-type patterns \cite{Louis87,Fernandez88,Herrmann89,Furukawa93}, such as similar to those seem in viscous fingering.
As is shown in Fig. \ref{m10s51g50L2+1.ps}, many cracks tend to grow simultaneously from the center toward the boundaries.
They are accompanied by a series of tip splittings and the total length of the cracks increases smoothly with contraction.
As $\tau$ becomes larger, it takes more time to complete the crack patterns because of the slower propagation of cracks.

\

\begin{figure}[t]
%
\begin{minipage}{7cm}
\epsfxsize=6cm \epsfbox{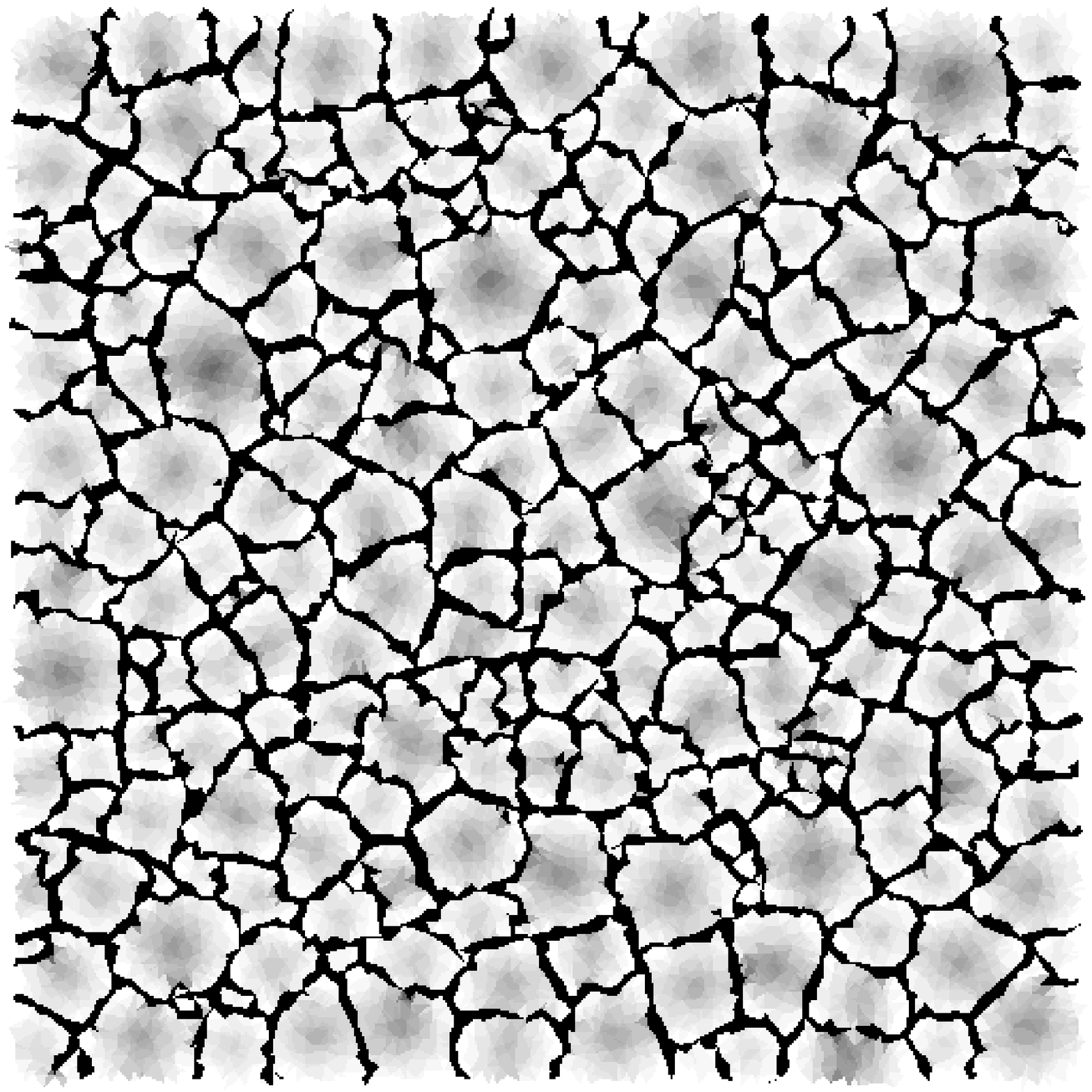}
\caption{A final crack pattern on a sticky bottom. $\hat{L}=20\protect\sqrt{2}$, $\hat{\mu}=1.0$, $\hat{\tau}=0.01$, $\hat{C}_s=1.0$ and $\hat{C}=10$.}
\label{m10s10g12L4+1.ps}
\end{minipage}\hspace{1cm}
%
\begin{minipage}{7cm}
\epsfxsize=6cm \epsfbox{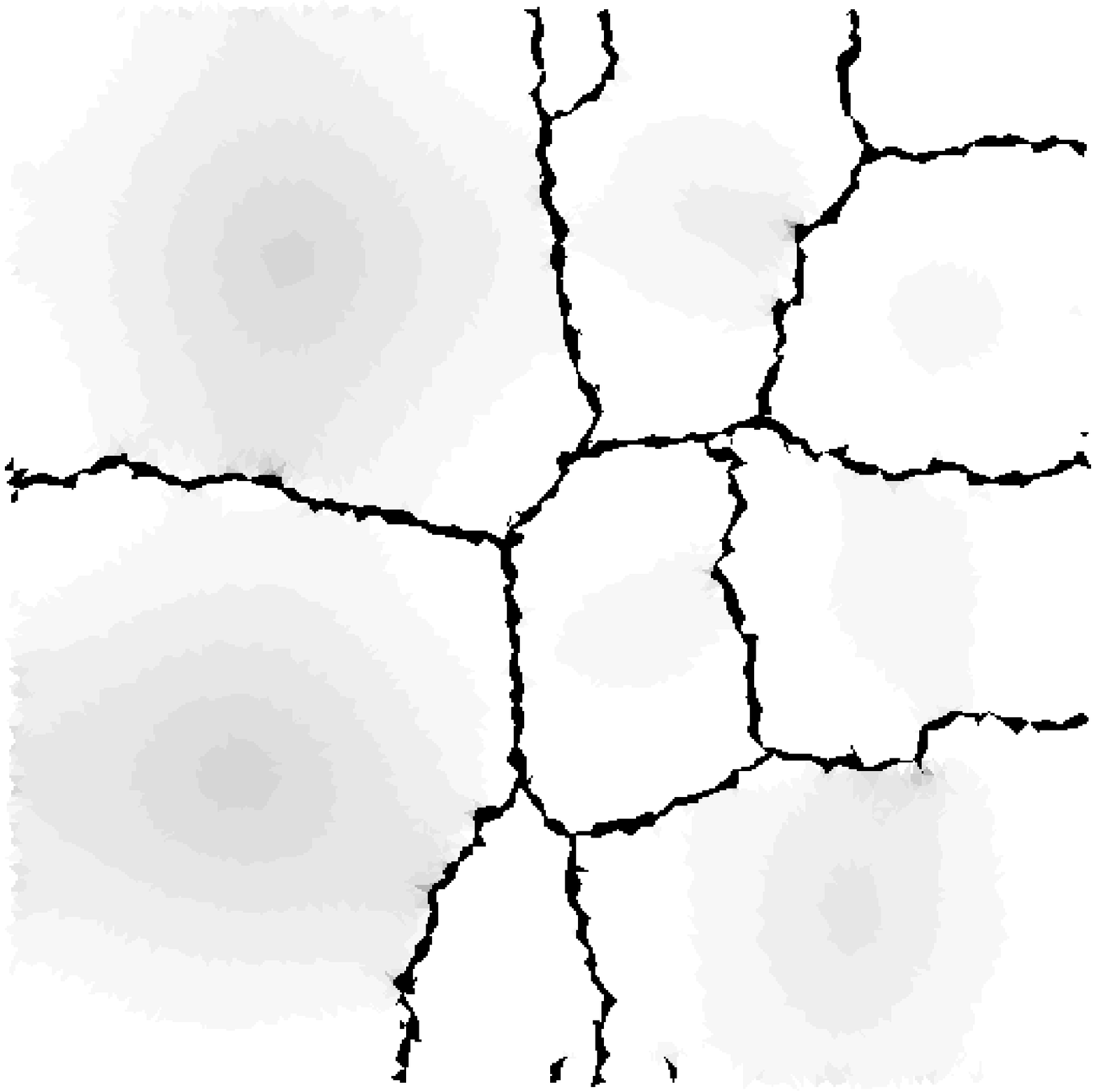}
\caption{A final crack pattern on a slippery bottom. $\hat{L}=20\protect\sqrt{2}$, $\hat{\mu}=1.0$, $\hat{\tau}=0.01$, $\hat{C}_s=0.1$ and $\hat{C}=10$.}
\label{m10s11g12L4+1.ps}
\end{minipage}

\vspace{0.5cm}
%
\begin{minipage}{7cm}
\epsfxsize=6cm \epsfbox{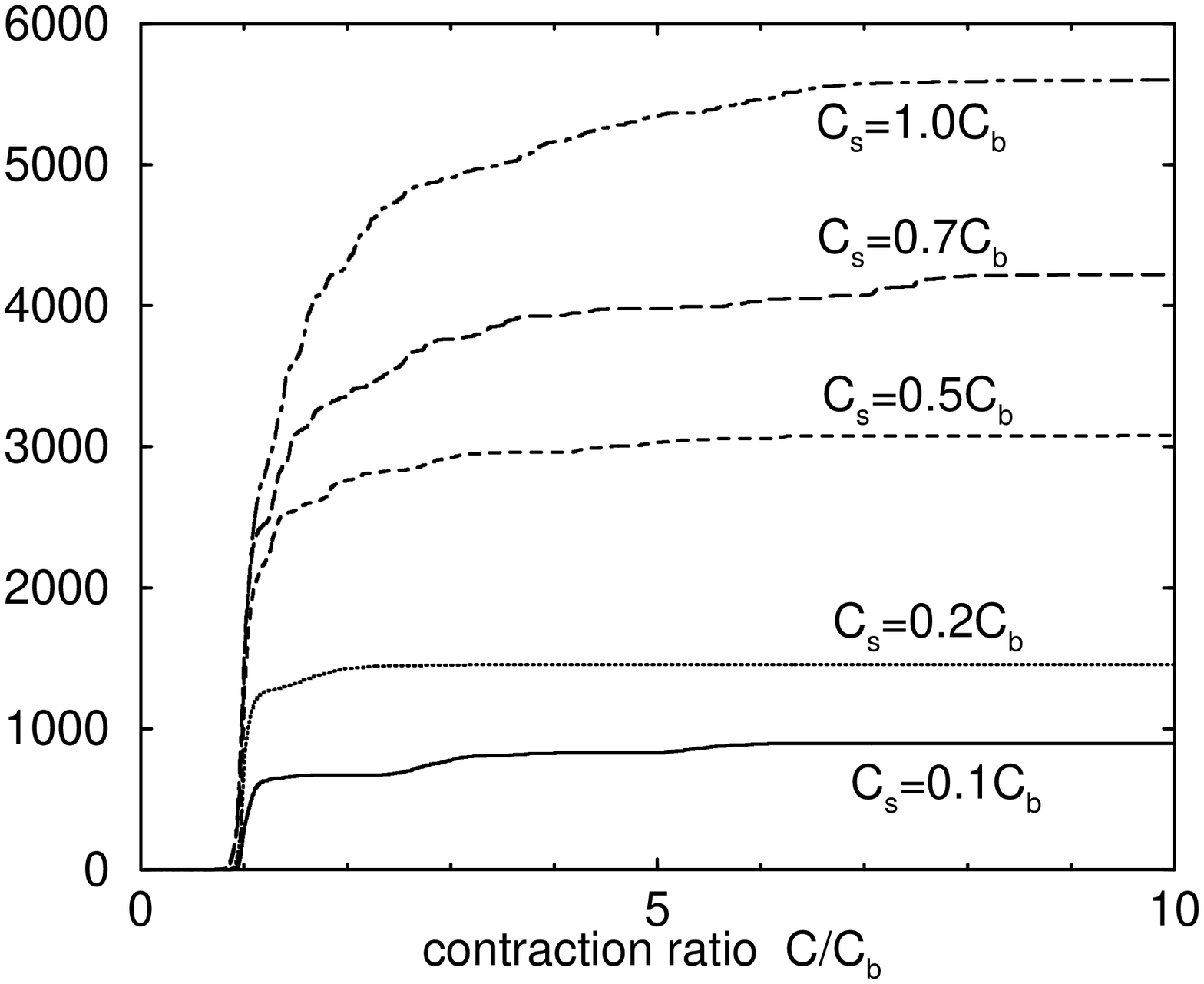}
\caption{The change of the total number of broken triangular cells for $\hat{L}=20\protect\sqrt{2}$, $\hat{\mu}=1.0$, $\hat{\tau}=0.01$ and $\hat{C}_s=0.1$, $0.2$, $0.5$, $0.7$, $1.0$.}
\label{m10g12L4+1.cellnum.eps}
\end{minipage}\hspace{1cm}
%
\begin{minipage}{7cm}
\epsfxsize=6cm \epsfbox{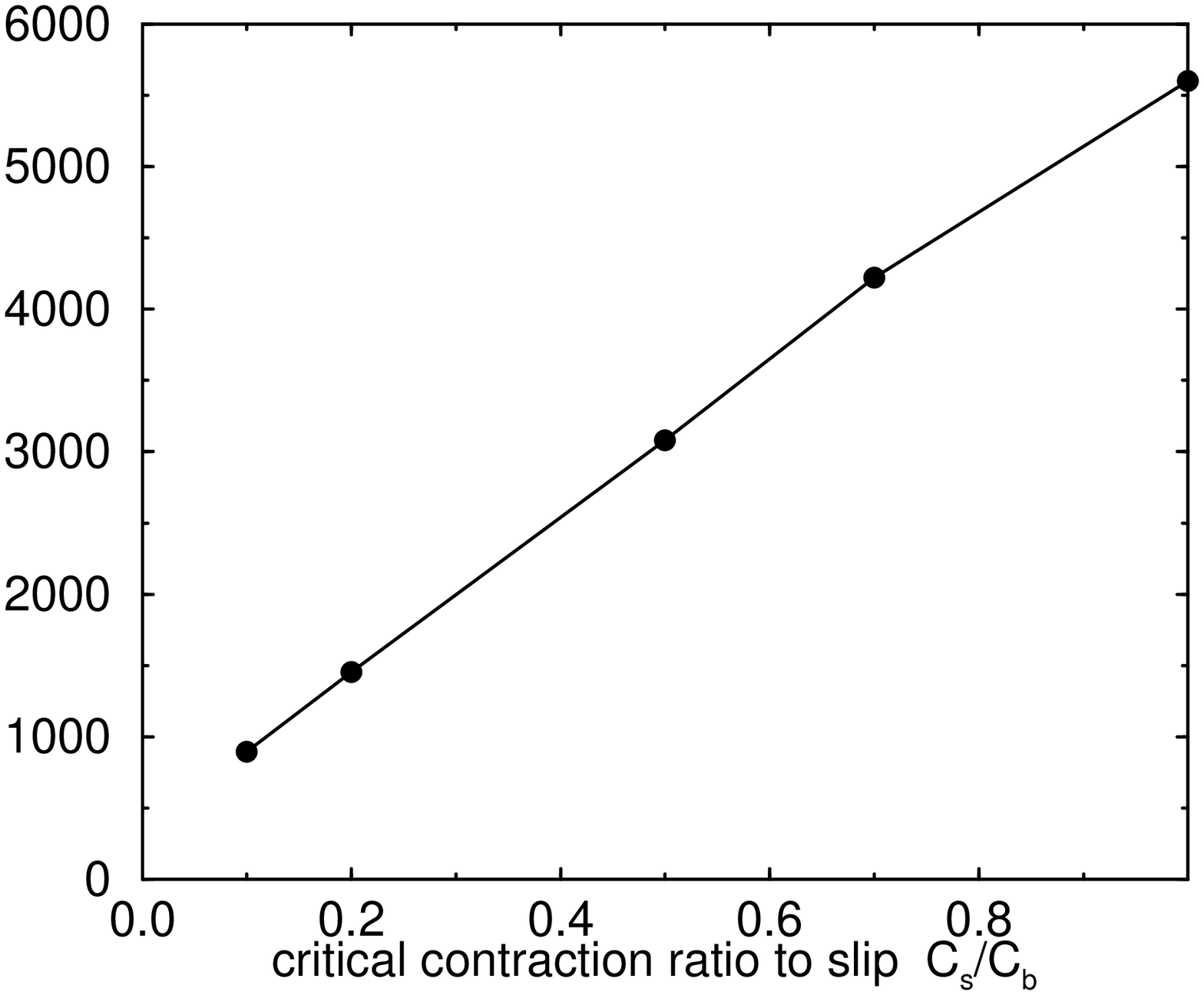}
\caption{The final number of broken triangles, where we plot the values at $\hat{C}=10$ in Fig. \protect\ref{m10g12L4+1.cellnum.eps} for $\hat{C}_s$.}
\label{m10g12L4+1.last_cellnum.eps}
\end{minipage}
\end{figure}

We can see the influence of slip on the crack patterns by changing $\hat{C}_s$ with the other parameters fixed. 
Figures \ref{m10s10g12L4+1.ps} and \ref{m10s11g12L4+1.ps} are snapshots of a crack pattern after full shrinking: $\hat{C}=10$ for $\hat{C}_s=1.0$ and $\hat{C}_s=0.1$, respectively.
Figure \ref{m10g12L4+1.cellnum.eps} shows the total number of broken triangular cells for the various values of $\hat{C}_s$.
This figure indicates that the crack patterns at $\hat{C}=10$ are close to final states.
As we expect, the final size of a cell becomes larger with smaller $\hat{C}_s$.
Figure \ref{m10g12L4+1.last_cellnum.eps} plots the total numbers of broken triangles at $\hat{C}=10$ for $\hat{C}_s$.
They increase monotonously in this range with an almost constant rate.

\

\begin{figure}[t]
%
\begin{minipage}{7cm}
\epsfxsize=6cm \epsfbox{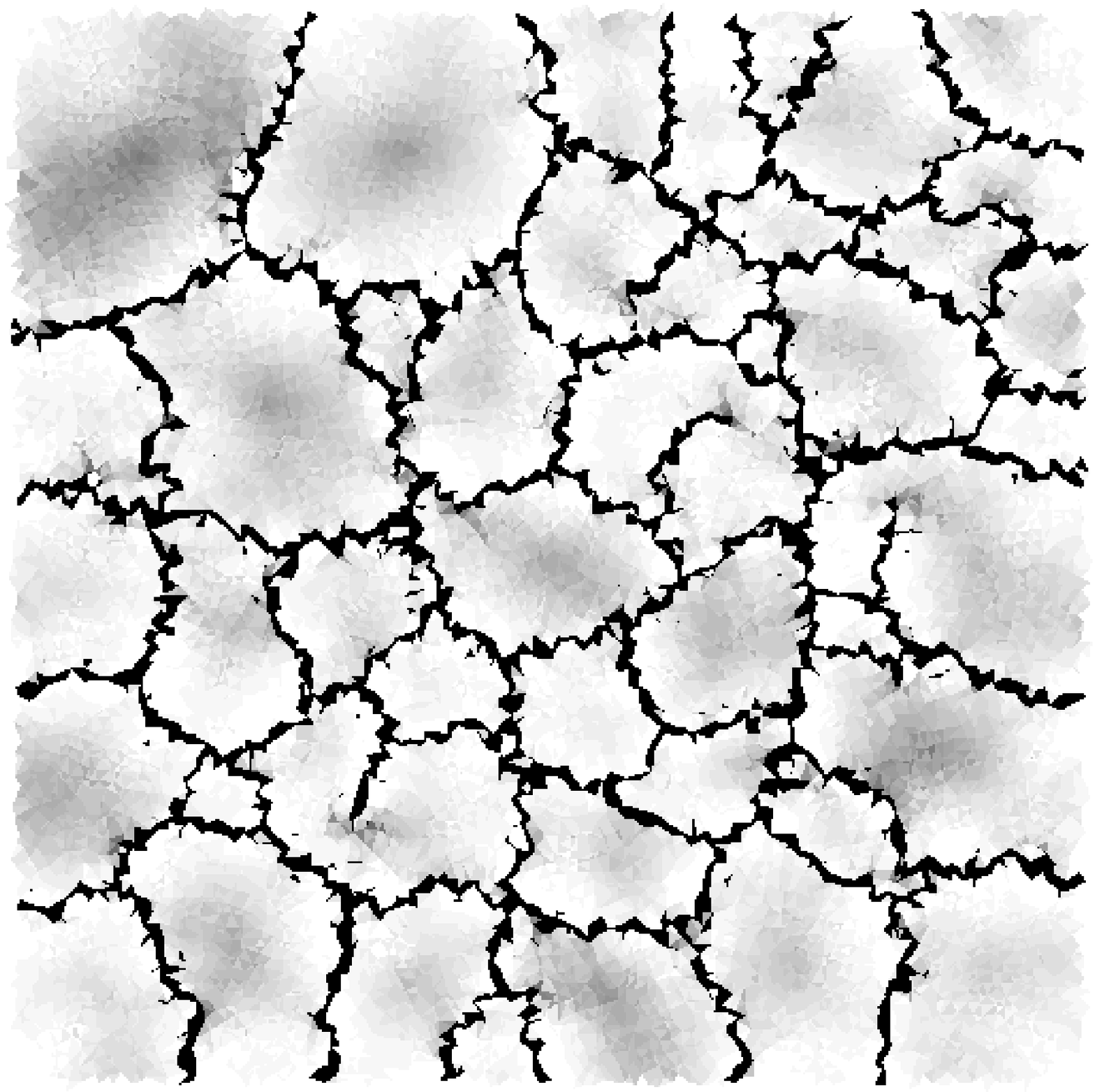}
\caption{A crack pattern for small $\hat{\mu}$. $\hat{L}=$$10\protect\sqrt{2}$, $\hat{\mu}=0.02$, $\hat{C}_s=0.5$, $\hat{\tau}=0.01$ and $\hat{C}=10$.}
\label{m22s51g12L2+1.ps}
\end{minipage}\hspace{1cm}
%
\begin{minipage}{7cm}
\epsfxsize=6cm \epsfbox{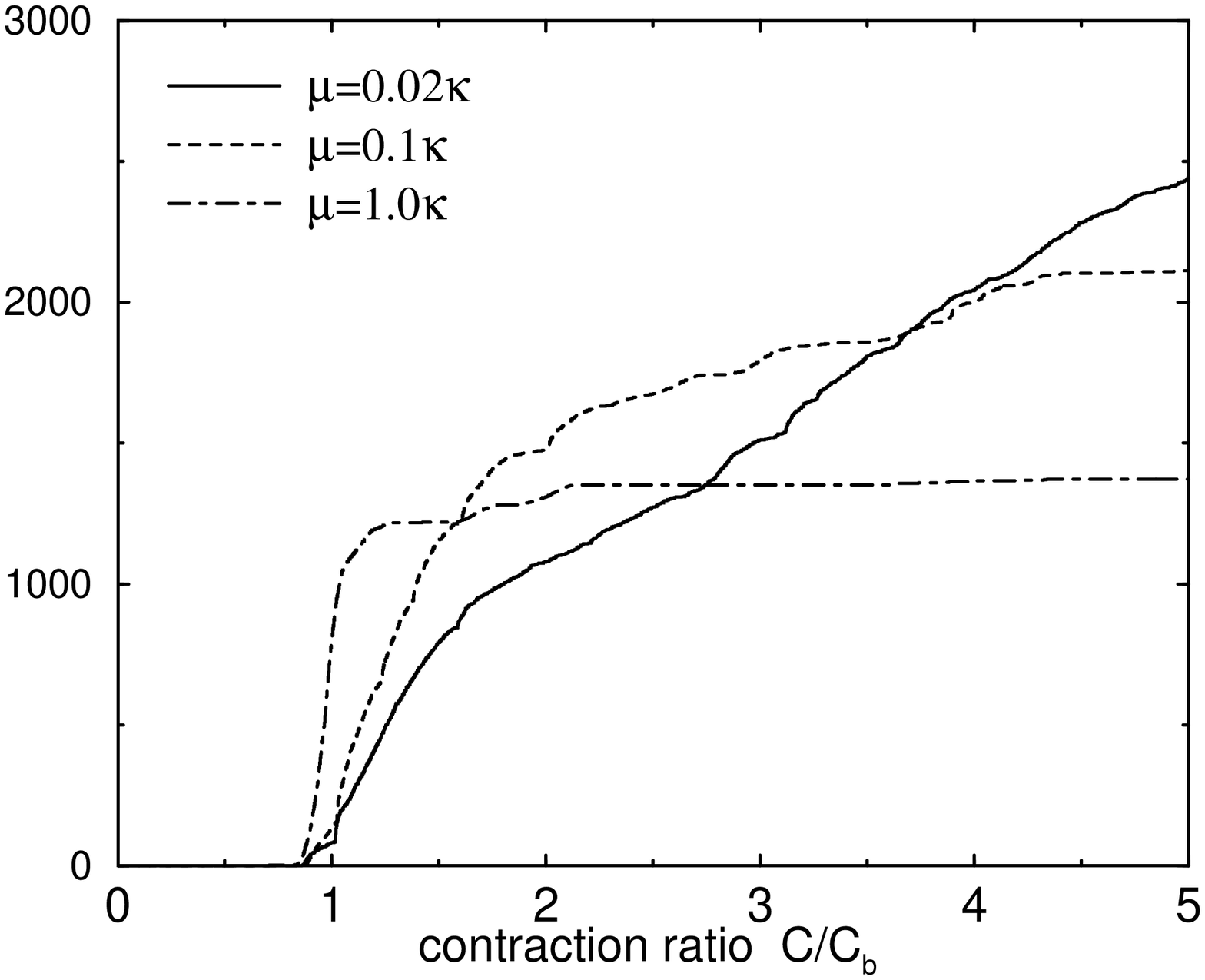}
\caption{The change of the total number of broken triangles for $\hat{L}=10\protect\sqrt{2}$, $\hat{C}_s=0.5$, $\hat{\tau}=0.01$ and $\hat{\mu}=0.02$, $0.1$, $1.0$. }
\label{s51g12L2+1.cellnum.eps}
\end{minipage}
\end{figure}

Next we change the elastic property with $\hat{\mu}$.
From the equations of $\sigma_{ij}$ (\ref{sigma yy 0}) and the crack speed (\ref{crack velocity}) in Sec. \ref{2d model-1}, we expect that, as $\hat{\mu}$ becomes smaller, the stress, excluding the dissipative force, has a weaker concentration at a crack tip, and the cracking speed is smaller.
Figure \ref{m22s51g12L2+1.ps} shows a snapshot of a crack pattern for $\hat{\mu}=0.02$.
We find that the cracks become irregular and jagged lines and that they propagate slowly.
The crack patterns also reach completion more and more slowly as $\hat{\mu}$ decreases, as is shown in Fig. \ref{s51g12L2+1.cellnum.eps}.

\begin{figure}[t]
%
\begin{minipage}{7cm}
\epsfxsize=6cm \epsfbox{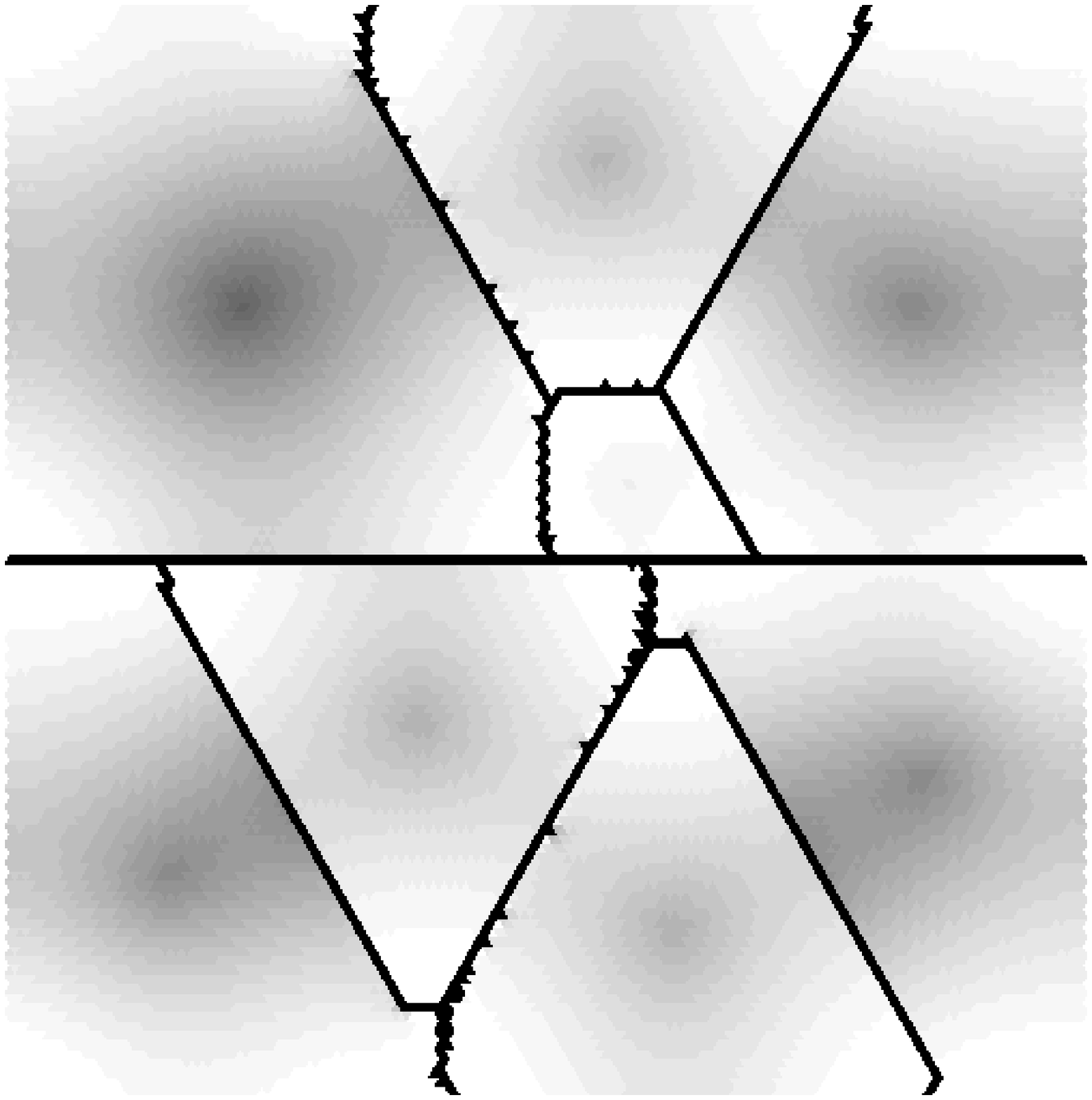}
\caption{A crack pattern on a regular triangular lattice for the same parameters as in Fig. \protect\ref{m10s51g12L2+1.ps} (d).}
\label{m10s51g12L2+1.regular.ps}
\end{minipage}\hspace{1cm}
\begin{minipage}{7cm}
\end{minipage}
\end{figure}

All of the above simulations were executed on the same random lattice.
For comparison, we also performed a simulation using a regular triangular lattice in the place of the random lattice.
Figure \ref{m10s51g12L2+1.regular.ps} shows the result using the same parameters used in Fig. \ref{m10s51g12L2+1.ps} except for the difference of the lattices.
It is clear that the anisotropy of the triangular lattice is reflected in the direction of cracks. 

\

Thus we can reproduce patterns similar to those of actual cracks using our two-dimensional model.
The dependence of the formation of cracks on the relaxation time and the elastic constants should be compared with actual experiments in more detail.
The experimental results of Groisman and Kaplan suggest a transition in the qualitative nature of patterns as the thickness is changed.
This is point (3) mentioned in Sec. \ref{Introduction}.
Similar changes of patterns are observed for slow relaxation or small $\hat{\mu}$ in our simulations.
However, our model does not contain the thickness $H$ explicitly because of the scaling with $H$, and we have no experimental data for the dependence of the other parameters on $H$.
In addition, it is possible that the inhomogeneity in a material plays an important role in this change because the size of a particle can become significantly large if $H$ is made sufficiently small\cite{Nishimoto99}. 
Obtaining a more detailed understanding that address these points is left as a future project.

\section{Conclusions}\label{Conclusions}

We studied the pattern formation of cracks induced by slow desiccation in a thin layer.
Assuming quasi-static and uniform contraction in the layer, we constructed a simple model in Sec. \ref{1d model-1}.
It models the layer as a linear elastic plane connected to elements on the bottom and considers the slip with a constant frictional force.

In Sec. \ref{1d model-2}, we considered the critical stress condition by introducing a critical value of the energy density.
This model explains the proportionality relation between the final size of a crack cell and the thickness of a layer and the experimental observations on the effect of slip.
We also considered the Griffith criterion as an alternative fragmentation condition.
However it predicts qualitatively different results for the final size of a crack.
Although these results need more detailed considerations, they suggest the possibility that dissipation in the bulk may be important in these materials.

In Sec. \ref{2d model-1}, we extended the model to two dimensions and introduced the relaxation of the elastic field to describe the development of cracks.
This is essentially the same as the Kelvin model for viscoelastic materials.
Because the stress excluding the dissipative force does not diverge at the tip of a propagating crack, we introduced a critical value as the breaking condition in front of a moving crack.
Assuming the existence of a stationary propagating crack, we obtained an estimation for the crack speed in closed form within a continuous theory.
This estimation explains the very slow propagation of actual cracks and predicts the proportionality relation between the crack speed and the thickness of a layer.
It is an open problem to understand the origin of the dissipation we introduced intuitively and the role of inhomogeneity in the stability of a crack.

In Sec. \ref{2d model-2}, we carried out numerical simulations of the two-dimensional model to investigate the formation of crack patterns.
By using the energy density defined on the microscopic cells of a lattice, we introduced a simplified breaking condition from the direct extension of the one-dimensional model.
We used a random lattice to remove the anisotropy of the lattice and obtained patterns similar to those observed in experiments.
We found that cracks grow in qualitative different ways depending on the ratio of the elastic constants and the relaxation time.
It is important in the connection with fingering patterns that for the slow relaxation crack patterns are formed by a succession of tip-splittings rather than by side-branching.
We need a better understanding of the role of inhomogeneity to explain the transition in the nature of patterns as a function of the thickness reported in the experiments.

For the experimental results of Groisman and Kaplan, which we mentioned in the points (1)-(3) in Sec. \ref{Introduction}, we believe the present results give qualitative explanations for (1) and (2) and a clue for (3), although more considerations is necessary.

\section*{Acknowledgements}

The paper of S. Sasa and T. Komatsu first motivated the author and is the starting point of this study.
The author had fruitful discussions with T. Mizuguchi, A. Nishimoto and Y. Yamazaki, who are also coworkers on project involving an experimental study of fracture.
The critical comments of S. Sasa and H. Nakanishi led to a reevaluation of the study.
G.C. Paquette are acknowledged for a conscientious reading of the manuscript.
Lastly the author is grateful to T. Uezu, S. Tasaki and the other members of our research group.


\pagebreak


\begin{figure}[h]
%
\begin{minipage}{7cm}
\epsfxsize=7cm
\epsfbox{figure/fig1.eps}
\caption{The one-dimensional model}
\label{1dmodel.eps}
\end{minipage}\hspace{1cm}\
%
\begin{minipage}{7cm}
\epsfxsize=7cm
\epsfbox{figure/fig2.eps}
\caption{The cross section of a shrinking cell of an elastic material.}
\label{boundary.eps}
\end{minipage}
\end{figure}

\begin{figure}[h]
%
\begin{minipage}{6cm}
\epsfxsize=6cm \epsfbox{figure/fig5.eps}
\caption{The two-dimensional model. The blank triangles represent a crack.}
\label{lattice.eps}
\end{minipage}\hspace{0.5cm}
%
\begin{minipage}{7cm}
\epsfxsize=6cm \epsfbox{figure/fig6.eps}
\caption{A part of a random lattice}
\label{Voronoi.eps}
\end{minipage}
\end{figure}

\begin{thebibliography}{99}
\nopagebreak
\bibitem{Kindle1917}
E. M. Kindle, Jour. Geol. {\bf 25}, 139 (1917).

\bibitem{Groisman94}
A. Groisman and E. Kaplan, Europhys. Lett. {\bf 25}, 415-20 (1994).

\bibitem{Ito98}
H. Ito and Y. Miyata, J. Geol. Soc. Japan     (in Japanese) {\bf 104}, 90-98 (1998).

\bibitem{Mitsui95}
Y. Mitsui, master thesis (in Japanese), Tohoku university, Japan (1995).

\bibitem{Nishimoto99}
A. Nishimoto, master thesis (in Japanese), Kyoto university, Japan (1999).

\bibitem{Allain95}
C. Allain and L. Limat, Phys. Rev. Lett. {\bf 74}, 2981-4 (1995).

\bibitem{Sasa97}
S. Sasa and T. Komatsu, Jpn. J. Appl. Phys. {\bf 6}, 391-5 (1997).

\bibitem{Hornig96}
T. Hornig, I. M. Sokolov and A. Blumen, Phys. Rev. E {\bf 54}, 4293-98 (1996).

\bibitem{Handge97}
U. A. Handge, I. M. Sokolov and A. Blumen, Europhys. Lett. {\bf 40}, 275-80 (1997).

\bibitem{Leung97}
K. Leung and J. V. Andersen, Europhys. Lett. {\bf 38}, 589-594 (1997).

\bibitem{Yuse93}
A. Yuse and M. Sano, Nature {\bf 362}, 329-31 (1993).

\bibitem{Yuse97}
A. Yuse and M. Sano, Physica D {\bf 108}, 365-78 (1997).

\bibitem{Ronsin97}
O. Ronsion and B. Perrin, Europhys. Lett. {\bf 38}, 435-40 (1997).

\bibitem{Hayakawa94a}
Y. Hayakawa, Phys. Rev. E {\bf 49}, R1804-7 (1994).

\bibitem{Hayakawa94b}
Y. Hayakawa, Phys. Rev. E {\bf 50}, R1748-51 (1994).

\bibitem{Sasa94}
S. Sasa, K. Sekimoto and H. Nakanishi, Phys. Rev. E {\bf 50}, R1733-6 (1994).

\bibitem{Holmes78}
A. Holmes, {\it Holmes Principles of Physical Geology} (John Wiley \& Sons,  New York, 1978).

\bibitem{Wearie83}
D. Weaire and C. O'Caroll, Nature {\bf 302}, 240-1 (1983).

\bibitem{Fineberg91}
J. Fineberg, S. P. Gross, M. Marder and H. L. Swinney, Phys. Rev. Lett. {\bf 67}, 457-460 (1991).

\bibitem{Marder96}
M. Marder, Nature {\bf 381}, 275-6 (1996).

\bibitem{Barber89}
M. Barber, J. Donley and J. S. Langer, Phys. Rev. A {\bf 40}, 366-376 (1989).

\bibitem{Langer92}
J. S. Langer, Phys. Rev. A {\bf 46}, 3123-31 (1992).

\bibitem{Langer93}
J. S. Langer and H. Nakanishi, Phys. Rev. E {\bf 48}, 439-48 (1993).

\bibitem{Ching96a}
E. S. S. Ching, J. S. Langer, and H. Nakanishi, Phys. Rev. E {\bf 53}, 2864-80 (1996).

\bibitem{Ching96b}
E. S. S. Ching, J. S. Langer, and H. Nakanishi, Phys. Rev. Lett. {\bf 76}, 1087-90 (1996).

\bibitem{Holian97}
B. L. Holian and R. Thomson, Phys. Rev. E {\bf 56}, 1071-9 (1997).

\bibitem{Fukuhara98}
T. Fukuhara and H. Nakanishi, J. Phys. Soc. Jpn. {\bf 67}, 4064-7 (1998).

\bibitem{Kessler98}
D. A. Kessler and H. Levine, preprint (1998).

\bibitem{Griffith1920}
A. A. Griffith, Philos. Trans. R. Soc. London  Ser.A {\bf 221}, 163-98 (1920).

\bibitem{LandauElasticTheory}
L. D. Landau and E. M. Lifshitz, {\it Theory of Elasticity, {\rm 3rd ed.}} (Butterworth Heinemann,  Oxford, 1995).

\bibitem{Lawn93}
B. Lawn, {\it Fracture of Brittle Solids} (Cambridge University Press,  New York, 1993).

\bibitem{Freund90}
L. B. Freund, {\it Dynamic Fracture Mechanics} (Cambridge University Press,  New York, 1990).

\bibitem{Friedberg84}
R. Friedberg and H. -C. Ren, Nuclear Physics B {\bf 235}, 310-20 (1984).

\bibitem{Moukarzel92}
C. Moukarzel, Physica A {\bf 190}, 13-23 (1992).

\bibitem{Louis87}
E. Louis and F. Guinea, Europhys. Lett. {\bf 3}, 871-7 (1987).

\bibitem{Fernandez88}
L. Fernandez, F.Guinea and E. Louis, J. Phys. A {\bf 21}, L301-5 (1988).

\bibitem{Herrmann89}
H. J. Herrmann, J. Kert\'{e}sz and L. de. Arcangelis, Europhys. Lett. {\bf 10}, 147-152 (1989).

\bibitem{Furukawa93}
H. Furukawa, Prog. Theor. Phys. {\bf 90}, 949-59 (1993).

\bibitem{Astrom98}
J. \AA str\"om, M. Alava and J. Timonen, Phys. Rev. E {\bf 57}, R1259-62 (1998).

\bibitem{Galdarelli98}
G. Galdarelli, R. Cafiero and A. Gabrielli, Phys. Rev. E {\bf 57}, 3878-85 (1998).

\bibitem{NumericalRecipes}
W. H. Press, B. P. Flannery, S. A. Teujiksky and W. T. Etterling, {\it Numerical Recipes} (Cambridge university press,  New York, 1997).
\end{thebibliography}
\end{document}